\documentclass[11pt,a4paper]{article}
\pdfoutput=1
\usepackage{jheppub}

\usepackage[english]{babel}
\usepackage{amsfonts,amsmath,amssymb,amsthm,bm,mathtools,slashed,empheq}
\usepackage{bbold}
\usepackage{comment,enumerate,footnote,graphicx,subfloat,relsize,footnote}
\usepackage{array,tabularx,tabu,multirow,framed,capt-of,colortbl}
\usepackage[usenames,dvipsnames]{xcolor}
\usepackage{booktabs,hhline}
\numberwithin{equation}{section}
\makeatletter
\g@addto@macro\bfseries{\boldmath}
\makeatother
\usepackage{cleveref}
\crefformat{pluralequation}{#2\black{eqs.~(}#1\black{)}#3}
\Crefformat{pluralequation}{#2\black{Equations~(}#1\black{)}#3}
\crefformat{pluralfigure}{#2\black{figs.~}#1#3}
\Crefformat{pluralfigure}{#2\black{Figures~}#1#3}
\usepackage{tikz}
\usetikzlibrary{arrows,shapes}
\usetikzlibrary{trees,patterns}
\usetikzlibrary{matrix,arrows} 				
\usetikzlibrary{positioning}				  
\usetikzlibrary{calc,through}				  
\usetikzlibrary{decorations.pathreplacing}  
\usepackage{pgffor}							

\usetikzlibrary{decorations.pathmorphing}	
\usetikzlibrary{decorations.markings}
\tikzset{
>=stealth', 
vector/.style={decorate, decoration={snake}, draw},
provector/.style={decorate, decoration={snake,amplitude=2.5pt}, draw},
antivector/.style={decorate, decoration={snake,amplitude=-2.5pt}, draw},
bigvector/.style={decorate, decoration={snake,amplitude=4pt}, draw},
fermion/.style={draw=black, postaction={decorate}, 
	decoration={markings,mark=at position .55 with {\arrow[draw=black]{>}}}},
fermionbar/.style={draw=black, postaction={decorate},
    decoration={markings,mark=at position .55 with {\arrow[draw=black]{<}}}},
fermionnoarrow/.style={draw=black},
doublefermion/.style={draw=black,double, postaction={decorate},
	decoration={markings,mark=at position .57 with {\arrow[draw=black]{>}}}},
doublefermionbar/.style={draw=black,double, postaction={decorate},
	decoration={markings,mark=at position .57 with {\arrow[draw=black]{<}}}},
doublefermionnoarrow/.style={draw=black,double},
gluon/.style={decorate, draw=black,
    decoration={coil,amplitude=4pt, segment length=5pt}},
scalar/.style={dashed,draw=black, postaction={decorate},
	decoration={markings,mark=at position .55 with {\arrow[draw=black]{>}}}},
scalarbar/.style={dashed,draw=black, postaction={decorate},
    decoration={markings,mark=at position .55 with {\arrow[draw=black]{<}}}},
scalarnoarrow/.style={dashed,draw=black},
momentum/.style={draw=black, postaction={decorate},
    decoration={markings,mark=at position 1 with {\arrow[draw=black]{>}}}},
antimomentum/.style={draw=black, postaction={decorate},
    decoration={markings,mark=at position 0.1 with {\arrow[draw=black]{<}}}}
}

\tikzstyle{block} = [draw, rectangle, minimum height=3em, minimum width=6em]
    
\newcommand{\nc}{\newcommand}
\nc{\pd}{\partial}
\nc{\bea}{\begin{eqnarray}}
\nc{\eea}{\end{eqnarray}}
\nc{\bal}{\begin{alignedat}}
\nc{\eal}{\end{alignedat}}
\nc{\beq}{\begin{equation}}
\nc{\eeq}{\end{equation}}
\nc{\bit}{\begin{itemize}}
\nc{\eit}{\end{itemize}}
\nc{\benu}{\begin{enumerate}}
\nc{\eenu}{\end{enumerate}}
\nc{\bdes}{\begin{description}}
\nc{\edes}{\end{description}}
\nc{\bma}{\begin{pmatrix}}
\nc{\ema}{\end{pmatrix}}

\newcommand{\black}[1]	{{\color{black} 	#1}}

\nc{\nn}{\nonumber}
\nc{\hc}{\text{h.c.}}
\nc{\cc}{\text{c.c.}}

\nc{\abs}[1]{\left| #1 \right|}
\def\[{\left[}
\def\]{\right]}
\def\({\left(}
\def\){\right)}
\def\<{\langle}
\def\>{\rangle}

\def\keV{{\rm keV}}

\def\GeV{{\rm GeV}}
\def\TeV{{\rm TeV}}

\def \cm{{\rm cm}}

\def\A			{\mathsmaller{A}}
\def\B			{\mathsmaller{B}}

\def\Bcal		{\mathsmaller{\cal B}}
\def\D			{\mathsmaller{D}}
\def\Dbar		{\mathsmaller{\bar{D}}}

\def\H			{\mathsmaller{H}}

\def\L			{\mathsmaller{L}}
\def\R			{\mathsmaller{R}}
\def\S			{\mathsmaller{S}}
\def\Scal		{\mathsmaller{\cal S}}
\def\T			{\mathsmaller{T}}

\def\W			{\mathsmaller{W}}

\def\aA			{\alpha_{\A}}
\def\aB			{\alpha_{\Bcal}}
\def\aR			{\alpha_{\R}}
\def\aS			{\alpha_{\Scal}}

\def\aH			{\alpha_{\H}}
\def\ann		{{\rm ann}}

\def\BW			{\B\W}
\def\BWH		{\B\W\H}

\def\BSF		{{\rm \mathsmaller{BSF}}}
\def\rBSF		{{\rm \mathsmaller{rBSF}}}

\def\dec		{{\rm dec}}
\def\DD			{\D\D}
\def\DDbar		{\D\Dbar}

\def\DS			{\D\S}

\def\DM			{\mathsmaller{\rm DM}}

\def\eff		{{\rm eff}}
\def\eq			{{\rm eq}}

\def\hH			{h_{\H}}
\def\gV			{g_{\mathsmaller{V}}}
\def\gW			{g_{\mathsmaller{W}}}
\def\gZ			{g_{\mathsmaller{Z}}}

\def\im			{\mathbb{i}}

\def\ion		{{\rm ion}}
\def\kappaB		{\kappa_{\Bcal}}

\def\mD			{m_{\D}}
\def\mH			{m_{\H}}
\def\mh			{m_{h}}
\def\mpl		{m_{\mathsmaller{\rm Pl}}}
\def\mS			{m_{\S}}

\def\mV			{m_{\mathsmaller{V}}}

\def\RB			{R_{\B}}
\def\RW			{R_{\W}}
\def\RH			{R_{\H}}

\def\SUL		{SU_\mathsmaller{L}(2)}

\def\SM			{\mathsmaller{\rm SM}}
\def\SS			{\S\S}
\def\SSDDbar	{\S\S\!\mathsmaller{/}\!\D\Dbar}
\def\trans		{{\rm trans}}

\def\UY			{U_\mathsmaller{Y}(1)}
\def\vH			{v_{\H}}
\def\vrel		{v_{\rm rel}}

\def\YD			{Y_{\D}}
\def\YDbar		{Y_{\Dbar}}
\def\YH			{Y_{\H}}
\def\YS			{Y_{\S}}

\def\zetaA		{\zeta_{\A}}

\def\zetaH		{\zeta_{\H}}
\def\zetaR		{\zeta_{\R}}
\def\zetaS		{\zeta_{\Scal}}

\begin{document}

\preprint{Nikhef-2021-004}
\title{Bound states of WIMP dark matter in Higgs-portal models. Part II. Thermal decoupling}

\author{Ruben Oncala and Kalliopi Petraki}

\affiliation{\href{http://www.lpthe.jussieu.fr/spip/index.php}{\color{black}
Sorbonne Universit\'e, CNRS, Laboratoire de Physique Th\'eorique et Hautes \'Energies, LPTHE, F-75005 Paris, France}}

\affiliation{
\href{https://www.nikhef.nl/en/}{\color{black} Nikhef}, 
Science Park 105, 1098 XG Amsterdam, The Netherlands}

\emailAdd{roncala@nikhef.nl}
\emailAdd{kpetraki@lpthe.jussieu.fr}

\abstract{
The Higgs \emph{doublet} can mediate a long-range interaction between multi-TeV particles coupled to the Weak interactions of the Standard Model, while its emission can lead to very rapid bound-state formation processes and bound-to-bound  transitions. Using the rates calculated in a companion paper, here we compute the thermal decoupling of multi-TeV WIMP dark matter coupled to the Higgs, and show that the formation of metastable dark matter bound states via Higgs-doublet emission and their decay decrease the relic density very significantly. This in turn implies that WIMP dark matter may be much heavier than previously anticipated, or conversely that for a given mass, the dark matter couplings to the Higgs may be much lower than previously predicted, thereby altering the dark matter phenomenology. While we focus on a minimal singlet-doublet model in the coannihilation regime, our calculations can be extended to larger multiplets where the effects under consideration are expected to be even more significant. 
}
\arxivnumber{2101.08667} 

\maketitle

\section{Introduction \label{Sec:Intro}}

Model-independent unitarity arguments~\cite{Baldes:2017gzw}, as well as model-specific calculations show that viable thermal-relic dark matter (DM) scenarios in the multi-TeV regime feature interactions that generate long-range potentials and give rise to bound states. The formation and decay of metastable bound states alters the DM decoupling in the early universe~\cite{vonHarling:2014kha} and contributes to its indirect signals~\cite{Pospelov:2008jd,MarchRussell:2008tu,An:2016kie,Cirelli:2016rnw,Asadi:2016ybp,Baldes:2017gzw,Baldes:2017gzu}. In fact, in a variety of models,  bound-state formation (BSF) can be faster than annihilation~\cite{Petraki:2015hla,Petraki:2016cnz,Harz:2018csl,Harz:2019rro,Oncala:2019yvj,Pearce:2013ola,Baldes:2020hwx}.  

This is particularly striking in models where DM couples to a light scalar boson charged under a symmetry. It was recently shown that the emission of such a scalar boson by a pair of interacting particles alters their effective Hamiltonian and can result in extremely rapid monopole capture processes~\cite{Oncala:2019yvj}. This may potentially have severe implications for multi-TeV DM coupled  to the Higgs doublet. Moreover, it has been shown that the 125~GeV Higgs boson can mediate a significant long-range interaction between TeV-scale particles, despite being heavier than all other Standard Model (SM) force mediators~\cite{Harz:2017dlj,Harz:2019rro}.

These considerations impel investigating the role of the Higgs doublet in the cosmology of multi-TeV DM coupled to the Higgs. The above effects can be potentially important in scenarios that involve a trilinear coupling between the DM and Higgs multiplets, 
i.e.~$\delta {\cal L} \supset \bar{\chi}_n H \chi_{n+1} +\hc$,
where $\chi_n$ stands for a fermionic or bosonic $n$-plet under $\SUL$, and DM is the lightest neutral mass eigenstate arising from the mixing of $\chi_n$ and $\chi_{n+1}$ after the electroweak phase transition (EWPT). These models have been among the archetypical scenarios of DM coupled to the Weak interactions of the SM (WIMPs)~\cite{Mahbubani:2005pt,DEramo:2007anh,Calibbi:2015nha,Banerjee:2016hsk,Cohen:2011ec,Cheung:2013dua,Fraser:2020dpy,Freitas:2015hsa,Lopez-Honorez:2017ora,Wang:2018lhk,Abe:2019wku,Dedes:2014hga,Yaguna:2015mva,Tait:2016qbg,Beneke:2016jpw,Bharucha:2018pfu,Filimonova:2018qdc,Betancur:2018xtj}.

Here we will focus on a minimal singlet-doublet realisation of this class of scenarios. In a companion paper, we have computed the non-relativistic potentials, the BSF cross-sections, the bound-state decay rates and bound-to-bound transitions~\cite{Oncala:2021tkz}. In this paper, we employ the results of~\cite{Oncala:2021tkz} to compute the DM thermal decoupling from the primordial plasma in the early universe. Several aspects of this calculation are examined in detail, including the effect of both radiative BSF and BSF via scattering on the relativistic thermal bath, the importance of bound-to-bound transitions and the late recoupling of the DM destruction processes due to BSF via Higgs emission. As in \cite{Oncala:2021tkz}, we carry out all computations assuming electroweak symmetry; indeed, for DM heavier than 5~TeV, freeze-out begins before the EWPT, even though the complete thermal decoupling may occur much later. The validity of the approximation is discussed in detail.

We note that our calculations are important broadly for scenarios that  introduce new stable species in the fashion considered here, even if these species do not account for DM. Besides specifying the parameters for which the relic density of the lightest new mass eigenstate matches the observed DM abundance, we show how the relic density is affected by the new effects within a broader parameter space. As is standard, the relic density of stable particles sets a cosmological constraint on new physics scenarios.

This work is organized as follows. In \cref{Sec:Model}, we summarise the model following \cite{Oncala:2021tkz}, and briefly review its basic properties in the broken electroweak phase. Moreover, we discuss the temperature dependence of the Higgs-doublet mass in relation to the BSF processes of interest. In \cref{Sec:FreezeOut}, we lay out the formalism for computing the DM thermal decoupling, and discuss the interplay among bound-state formation, ionisation, decay and bound-to-bound transition processes. The results of this computation are presented and discussed in \cref{Sec:FreezeOut_Results}. We conclude in \cref{Sec:Conclusion}.

\clearpage
\section{The model \label{Sec:Model}}

\subsection{Lagrangian and mass eigenstates \label{sec:Model_Lagrangian}}

We begin by specifying the model following ref.~\cite{Oncala:2021tkz}, before summarising the mass eigenstates and interactions in the broken electroweak phase. 

We consider a gauge-singlet Majorana fermion
$S = (\psi_\alpha, \psi^{\dagger \dot{\alpha}})^{\T}$ 
of mass $\mS$, and a Dirac fermion  
$D=(\xi_{\alpha}, \chi^{\dagger \dot{\alpha}})^{\T}$ 
of mass $\mD$ with SM gauge charges $\SUL \times \UY=(\mathbb{2}, 1/2)$. $S$ and $D$ are assumed to be odd under a $\mathbb{Z}_2$ symmetry that leaves all the SM particles unaffected. Under these assignments, the new degrees of freedom (dof) allow to extend the SM Lagrangian by the following interactions
\begin{align}
{\delta \cal L} &= 
 \frac{1}{2} \bar{S} (\im \slashed{\partial} -\mS )S
+\overline{D} (\im \slashed{\cal D} -\mD )D
-(y_{\L} \bar{D}_{\L} H S + y_{\R} \bar{D}_{\R} H S + \hc),
\label{eq:Lagrangian}
\end{align}
where $H$ is the SM Higgs doublet of mass $\mH$ and hypercharge $\YH = 1/2$, and 
$D_{\L} \equiv P_{\L} D = (\xi_{\alpha},0)^{\T}$ and $D_{\R} \equiv P_{\R} D = (0,\chi^{\dagger \dot{\alpha}})^{\T}$, with $P_{\R,\L} = (1 \pm \gamma_5)/2$ being the right-handed and left-handed projection operators. In the above, ${\cal D}_\mu \equiv \partial_\mu - \im g_1 Y B_\mu - \im g_2 W_\mu^a t^a$ is the covariant derivative, with $t^a = \frac{1}{2} (\sigma^1, \sigma^2, \sigma^3)$ and $\sigma$ being the Pauli matrices. The particle content of \cref{eq:Lagrangian} is summarised in \cref{tab:particles}.
\begin{table}[h!]
\centering
\begin{equation*}
\begin{tabular}{|c|c|c|c|}
\hline
field & $\SUL$ & $\UY$ & $\mathbb{Z}_2$ \\
\hline
$S$ & 1 & 0   & $-1$\\
$D$ & 2 & 1/2 & $-1$\\
$H$ & 2 & 1/2 & $+1$\\
\hline
\end{tabular}
\end{equation*}
\caption{\label{tab:particles} Particle content and charge assignments}
\end{table}

We take $\mS$ and $\mD$ to be real. This can be achieved by rephasing $\psi$ and either $\xi$ or $\chi$. Rephasing the remaining spinor eliminates the phase of one of the Yukawa couplings. Thus the free parameters of the present model are 4 real couplings (two masses and two dimensionlesss Yukawa couplings), and a phase that allows for CP violation.

We will focus on the regime where $S$ and $D$ can co-annihilate efficiently before the EWPT of the universe. This occurs if their masses are similar, within about $10\%$. 
A larger mass difference would imply that the number density of the heavier species in the non-relativistic regime becomes negligible, since it would be suppressed with respect to the lighter species by $\exp[-(\delta m/m) x] \lesssim 0.1$, with $x\equiv m/T \sim 25$ during DM freeze-out. 
Since the masses must be similar for the processes considered here to be relevant, we take them to be equal for simplicity,
\begin{align}
\mD = \mS \equiv m .
\label{eq:MassEquality}
\end{align}
It will be useful also to introduce the reduced mass of a pair of DM particles,
\begin{align}
\mu \equiv m/2 .
\label{eq:ReducedMass}
\end{align}

In addition, in order to reduce the number of free parameters, we set
\begin{align}
y_{\L} = y_{\R} \equiv y ,
\label{eq:YukawaEquality}
\end{align}
which we take to be real. (The CP violation is not important for our purposes.) Our computations can of course be extended to more general Yukawa couplings. As is standard, we define the couplings
\begin{align}
\alpha_1 \equiv \frac{g_1^2}{4\pi}, \qquad
\alpha_2 \equiv \frac{g_2^2}{4\pi}, \qquad
\aH \equiv \frac{y^2}{4\pi}.
\label{eq:alphas_def}
\end{align}
For later convenience, following~\cite{Oncala:2021tkz}, we also define the couplings
\begin{subequations}
\label{eq:GaugeSinglet_alphas_def}
\label[pluralequation]{eqs:GaugeSinglet_alphas_def}
\begin{align}
\aR &\equiv \frac{1}{2} \[ \sqrt{ [(\alpha_1+3\alpha_2)/4]^2 + 8\aH^2 } - (\alpha_1+3\alpha_2)/4 \] ,
\label{eq:GaugeSinglet_alphaR_def} 
\\
\aA &\equiv \frac{1}{2} \[ \sqrt{ [(\alpha_1+3\alpha_2)/4]^2 + 8\aH^2 } + (\alpha_1+3\alpha_2)/4 \] .
\label{eq:GaugeSinglet_alphaA_def}
\end{align}
\end{subequations}

The model of \cref{eq:Lagrangian} and various aspects of its phenomenology have been considered for general parameters extensively in the past~\cite{Mahbubani:2005pt,DEramo:2007anh,Cohen:2011ec,Cheung:2013dua,Calibbi:2015nha,Banerjee:2016hsk,Fraser:2020dpy,Freitas:2015hsa,Lopez-Honorez:2017ora,Wang:2018lhk}. Here we will only briefly review the mass eigenstates and their interactions after electroweak symmetry breaking for the choice of parameters denoted in \cref{eq:MassEquality,eq:YukawaEquality}.

\subsection*{Mass eigenstates in the broken electroweak phase}

At electroweak symmetry breaking, the neutral component of the Higgs doublet acquires a vacuum expectation value. In terms of their $\SUL$ components, the $H$, $\xi$ and $\chi$ fields are
\begin{align}
H = \(
\begin{array}{c} \phi^+ \\ \dfrac{1}{\sqrt{2}} (\vH + h + \im \phi^0) \end{array}
\), \qquad
\xi_\alpha = \(
\begin{array}{c} \xi^+_{\alpha} \\ \xi^0_{\alpha} \end{array}
\), \qquad
\chi_\alpha = \(
\begin{array}{c} \chi^0_{\alpha} \\ \chi^-_{\alpha} \end{array}
\), 
\label{eq:SU2Lcomponents}
\end{align} 
with $\vH \simeq 246~\GeV$ being the Higgs vacuum expectation value.

We define the left-handed multiplet of the neutral states
$\hat{N}_\alpha \equiv (\psi_\alpha, \xi_\alpha, \chi_\alpha)^{\T}$. 
Then, by inserting \cref{eq:SU2Lcomponents} into the Lagrangian \eqref{eq:Lagrangian}, we find the corresponding mass terms,
\begin{align}
\delta {\cal L}_{\mathsmaller{N},\rm mass} = 
- \frac{1}{2} \hat{N}^\alpha \hat{M}_{\mathsmaller{N}} \hat{N}_\alpha +\hc ,
\label{eq:Lagrangian_MassNeutral} 
\end{align}
with
\begin{align}
\hat{M}_{\mathsmaller{N}} = \(
\begin{array}{ccc}
m ~&~ y\vH / \sqrt{2} ~&~ y\vH / \sqrt{2} \\
y\vH / \sqrt{2} ~&~ 0 ~&~ m \\
y\vH / \sqrt{2} ~&~ m ~&~ 0
\end{array}
\).
\label{eq:MassNeutral} 
\end{align}
We diagonalise \cref{eq:Lagrangian_MassNeutral} by setting
\begin{align}
N_\alpha = U \hat{N}_\alpha, \qquad
N^\alpha = \hat{N}^\alpha U^{\T}, \qquad
M_{\mathsmaller{N}} = (U^{\T})^{-1} \hat{M}_{\mathsmaller{N}} U^{-1},
\end{align}
where $U$ is the unitary matrix
\begin{align}
U = \(
\begin{array} {ccc}
-1/\sqrt{2}	~~&~~1/2 			~~&~~ 1/2 \\
0 			~~&~~\im/\sqrt{2}	~~&~~ -\im/\sqrt{2} \\
 1/\sqrt{2}	~~&~~1/2 			~~&~~ 1/2
\end{array}
\).
\label{eq:DiagonalisationMatrix}
\end{align}
The corresponding mass eigenvalues are
\begin{align}
m_1 \equiv m-y\vH, \qquad
m_2 \equiv m, \qquad
m_3 \equiv m+y\vH.
\label{eq:MassEigenvalues}
\end{align}

In addition to the neutral states, there is a charged Dirac fermion of mass $m$.

\subsection*{Interactions in the broken electroweak phase and constraints}

The interactions among the neutral states, $N\equiv (N_1,N_2,N_3)^{\T}$, are described by the Lagrangian~\cite{Calibbi:2015nha}
\begin{align}
\delta {\cal L}_{\mathsmaller{N},\rm inter} = 
\frac{g_2}{2c_{\mathsmaller{W}}} Z_\mu N_2^\dagger \bar{\sigma}^\mu 
\frac{\im }{2\sqrt{2}} \( N_1 + N_3 \)
-\dfrac{y}{2} h \( 
N_1 N_1 - N_3 N_3 + N_1 N_3
\) +\hc,
\end{align}
where $c_{\mathsmaller{W}} = g_2/{\sqrt{g_1^2 + g_2^2}}$.

Since the coupling to $Z_\mu$ is non-diagonal, with the mass splitting being always much larger than $\sim{\cal O}(100~\keV)$ for the $y$ values we will consider here (cf.~\cref{Sec:FreezeOut}), the constraints from direct detection experiments due to this interaction are evaded. On the other hand, the coupling to the Higgs boson is expected to yield sizable DM-nucleus scattering and potentially strong constraints. Existing analyses of the direct detection data for this model do not extend to the multi-TeV regime that is of interest here. Moreover, the direct detection constraints on the model \eqref{eq:Lagrangian} are generally significantly relaxed around the so-called blind spots where the coupling to the Higgs vanishes, roughly when $y_{\mathsmaller{L}} =-y_{\mathsmaller{R}}$~(see e.g.\cite{Cheung:2013dua,Calibbi:2015nha}.) 
Constraints may also arise from electroweak precision observables, and in particular from the contribution to the $T$ parameter. While the latter scales as $(y_{\L}^2 - y_{\R}^2)^2$ and vanishes in the limit considered here, it may become important for large values of the Yukawa coupling(s) if $|y_{\L}| \gg |y_{\R}|$ or $|y_{\L}| \ll |y_{\R}|$. We refer to~\cite{DEramo:2007anh,Calibbi:2015nha,Banerjee:2016hsk} for related studies.

While a detailed phenomenological analysis is beyond the scope of the present work, our results are important for interpreting the experimental constraints, since they imply a different relation between the DM mass and couplings in order for the observed DM density to be attained via thermal freeze-out.

\subsection{Bound states \label{sec:Model_BoundStates}}

The interactions of \cref{eq:Lagrangian} -- in particular the exchange of $B$, $W$ and $H$ bosons -- generate long-range potentials among the $S$, $D$ and $\bar{D}$ species that distort the wavefunctions of scattering states and give rise to bound states. The long-range dynamics in the symmetric electroweak phase has been discussed in detail in the companion paper~\cite{Oncala:2021tkz}, and shall not be repeated here. However, since our focus is the effect of bound states on the DM thermal decoupling, in \cref{tab:BoundStatesGroundLevel} we summarise for convenience the bound levels we consider.

\begin{table}[h!]
\centering
\renewcommand*{\arraystretch}{1.2}
\begin{align*}
\begin{array}{ |c|c|c|c|c|c| } 
\hline
\text{Bound state } {\cal B}
&\UY	
&\SUL
&\text{Spin}	
&\text{dof } (g_{\B})	
&\text{Bohr momentum } (\kappaB)
\\ \hline \hline
SS/D\bar{D}	
&0
&\mathbb{1}
&0
&1
&m\aA / 2
\\ \hline
D\bar{D}	
&0	
&\mathbb{1}	
&1
&3
&m (\alpha_1+3\alpha_2) / 8 
\\ \hline
DD	
&1	
&\mathbb{1}	
&1
&3
&m(-\alpha_1+3\alpha_2) / 8
\\ \hline
DS	
&1/2	
&\mathbb{2}	
&0
&2
&m\aH / 2
\\  \hline
\end{array}
\end{align*}
\caption{\label{tab:BoundStatesGroundLevel} 
The ground-level (principal and angular-momentum quantum numbers $\{n\ell m\} = \{100\}$) bound states and their Bohr momenta $\kappaB$ in the limit $\mH \to 0$. The binding energies are $|{\cal E}_{\B}| = \kappaB^2/m$. The couplings are defined in \cref{eq:alphas_def,eq:GaugeSinglet_alphas_def}.
The $SS/D\bar{D}$, $D\bar{D}$ and $DD$ bound states can decay directly into radiation. The $DS$ rate of decay into radiation is suppressed, however the $DS$ bound state can transit spontaneously into an $SS/D\bar{D}$ bound state via $H$ emission. All other bound-to-bound transitions are suppressed. The bound state decay and transition rates are summarised in \cite[table~6]{Oncala:2021tkz}.}
\end{table}

\subsection{Higgs doublet mass and EWPT}

The cross-sections for BSF via $H$ emission depend on the Higgs doublet mass~\cite{Oncala:2021tkz}. Taking into account the finite temperature 1-loop corrections to the effective potential (see e.g.~\cite{Carrington:1991hz,Ahriche:2007jp}), we estimate that before the EWPT of the universe, the Higgs doublet mass was
\begin{align}
\mH^2 (T) \approx -\frac{\mh^2}{2} 
+ \frac{\pi T^2}{4} \(\alpha_1 + 3\alpha_2 + \frac{2\lambda}{\pi} +\frac{y_t^2}{\pi} \),
\label{eq:mH}
\end{align}
where $\mh \simeq 125~\GeV$ is the Higgs boson mass at zero temperature, $\lambda = \mh^2/(2\vH^2) \simeq 0.13$ is the Higgs quartic coupling, $V_{\SM} \supset -\lambda |H|^4$, and $y_t \simeq 0.994$ is the top quark Yukawa coupling. The EWPT occurs as $\mH^2 (T) \to 0$, i.e. at estimated temperature
\begin{align}
T_{\mathsmaller{\rm EWPT}} \approx 
\frac{\sqrt{2} \mh}{\sqrt{\pi \alpha_1 + 3\pi \alpha_2 +2\lambda +y_t^2}}
\simeq 151~\GeV .
\label{eq:T_EWPT}
\end{align}
In computing the DM decoupling, we use \cref{eq:mH} at $T \geqslant T_{\mathsmaller{\rm EWPT}}$, and set $\mH \to \mh$ at $T < T_{\mathsmaller{\rm EWPT}}$ while still using the annihilation and BSF rates computed under the assumption of electroweak symmetry. We discuss this approximation in \cref{sec:FreezeOut_Approximations}.

We may now estimate whether or when $\mH(T)$ implies that BSF via Higgs emission is kinematically suppressed. In a thermal distribution, the energy dissipated during BSF averages to 
$\<\omega \> = 3T/2 + |{\cal E}_{\Bcal}|$ (cf.~\cref{eq:omega} and \cite{Oncala:2021tkz}.) 
The first term suffices to provide for $\mH(T)$ for all $T>T_{\mathsmaller{\rm EWPT}}$ since $\mH (T > T_{\mathsmaller{\rm EWPT}}) \lesssim 0.63 T$, as well as after the EWPT, down to temperatures $T \sim 2\mh/3 \simeq 83~\GeV$. However, since the BSF cross-sections weigh preferentially low values of $\vrel$, the kinematic suppression may become important at somewhat larger $T$ than this estimate implies, unless $|{\cal E}_{\Bcal}|$ is sufficient to provide for $\mH$.

\clearpage
\section{Boltzmann equations for dark matter thermal decoupling \label{Sec:FreezeOut}}

\subsection{Coupled Boltzmann equations \label{sec:FreezeOut_Boltzmann}}

Let $Y_j \equiv n_j/s$ and  $Y_{\Bcal} \equiv n_{\Bcal}/s$ be the number-density-to-entropy-density ratios of the free species $j$ and the bound state ${\cal B}$ respectively. In our model, $j=S$, $D$, $\bar{D}$ and ${\cal B} = SS/D\bar{D}$, $D\bar{D}$, $DD$, $DS$ (cf.~\cref{tab:BoundStatesGroundLevel}.) We are ultimately interested in the total DM yield
\begin{align}
Y \equiv \YS + \YD + \YDbar = \YS + 2 \YD .
\label{eq:DensityTotal_def}
\end{align}
Note that the bound states are metastable and their abundance becomes eventually negligible, so we do not include them in \cref{eq:DensityTotal_def}. As is standard, we will use the time parameter
\begin{align}
x \equiv m/T .
\label{eq:x_def}
\end{align}
The entropy density of the universe is 
$s= (2\pi^2/45) g_{*\mathsmaller{S}} T^3
  = (2\pi^2/45) g_{*\mathsmaller{S}} m^3/x^3$.  
We denote by $g_{*\mathsmaller{S}}$ and $g_*$ the entropy and energy dof respectively, and define 
\begin{align}
g_{*,\eff}^{1/2} &= 
\frac{g_{*\mathsmaller{S}}}{\sqrt{g_*}}
\(1-\frac{x}{3g_{*\mathsmaller{S}}} \frac{dg_{*\mathsmaller{S}}}{dx} \) .
\label{eq:gstareff}
\end{align}
The evolution of $Y_j$ and $Y_{\Bcal}$ is governed by the coupled Boltzmann equations
\begin{subequations}
\label{eq:BoltzmannEqs}
\label[pluralequation]{eqs:BoltzmannEqs}
\begin{align}
\frac{dY_j}{dx} =& 
- \frac{\lambda}{x^2} \sum_i \<\sigma_{ji}^{\ann} \vrel\> \(Y_j Y_i - Y_j^{\eq}Y_i^{\eq}\)  
- \frac{\lambda}{x^2} \sum_i \sum_{\Bcal}\<\sigma_{j i\to \Bcal}^{\BSF} \, \vrel\> 
\(Y_j Y_i - \frac{Y_\Bcal}{Y_{\Bcal}^{\eq}} Y^{\eq}_j Y^{\eq}_i\) 
\nn \\  & 
- \Lambda \, x  \sum_{i} \<\Gamma_{j\to i}\> 
\(Y_j - \dfrac{Y_i}{Y_i^{\eq}} Y_j^{\eq}\) ,
\label{eq:BoltzmannEqs_Free} 
\\
\frac{dY_\Bcal}{dx} =& 
- \Lambda \, x \[ 
\< \Gamma_{\Bcal}^{\dec} \>   \(Y_{\Bcal} - Y_{\Bcal}^{\eq} \)
+\sum_{i,j} \< \Gamma_{\Bcal \to ij}^{\ion} \>
\(Y_{\Bcal} -  \dfrac{Y_i Y_j}{Y_i^{\eq} Y_j^{\eq}} Y_{\Bcal}^{\eq}\)
\right. \nn \\ &\left. 
+ \sum_{\Bcal' \neq \Bcal} 
\<\Gamma_{\Bcal\to \Bcal'}^{\trans} \>
\(Y_{\Bcal} - \dfrac{Y_{\Bcal'}}{Y_{\Bcal'}^{\eq}} Y_{\Bcal}^{\eq} \)
\] ,
\label{eq:BoltzmannEqs_Bound} 
\end{align}
\end{subequations}
where 
\begin{align}
\lambda \equiv 
\sqrt{\frac{\pi}{45}}\mpl m g_{*,\text{eff}}^{1/2}
\qquad \text{and} \qquad
\Lambda \equiv \dfrac{\lambda}{s \, x^3}
=\sqrt{\frac{45}{4\pi^3}} \frac{\mpl}{m^2} \frac{g_{*,\eff}^{1/2}}{g_{*,\S}} ,
\label{eq:lambdaAndLambda}
\end{align}
and the equilibrium densities in the non-relativistic regime are
\begin{align}
Y_i^\eq \simeq \frac{90}{(2\pi)^{7/2}} 
\frac{g_i}{g_{*,\mathsmaller{S}}} 
\, x^{3/2} \, e^{-x} 
\qquad \text{and} \qquad
Y_{\Bcal}^\eq \simeq \frac{90}{(2\pi)^{7/2}} 
\frac{g_{\Bcal}}{g_{*,\mathsmaller{S}}} 
\, (2x)^{3/2} \, e^{-2x} \, e^{|{\cal E}_{\Bcal}|/T} ,
\label{eq:Yeq_B}
\end{align}
where $g_i$ are the spin and $\SUL$ dof of the free species, with $g_\S = 2$, $g_\D = g_\Dbar = 4$. The dof $g_\Bcal$ and the binding energies ${\cal E}_{\Bcal}$ of the bound states we consider are listed in \cref{tab:BoundStatesGroundLevel}. For later convenience, we also define the total DM dof $g_{\DM} \equiv g_{\S}+g_{\D}+g_{\Dbar}=10$, and the equilibrium density of \eqref{eq:DensityTotal_def}
\begin{align}
Y^\eq &= \frac{90}{(2\pi)^{7/2}} 
\frac{g_\DM}{g_{*,\mathsmaller{S}}} 
\, x^{3/2} \, e^{-x} .
\label{eq:Yeq}
\end{align}

In the above, $\Gamma_{\Bcal}^{\dec}$, $\Gamma_{\Bcal\to ij}^{\ion}$ and $\Gamma_{\Bcal\to\Bcal'}^{\trans}$ are respectively the rates of ${\cal B}$ decay into radiation, ionisation (a.k.a.~dissociation) to $ij$, and transition into the bound level ${\cal B}'$. The rates $\Gamma_{j\to i}$ describe the transitions between free particles, due to decays, inverse decays and/or scatterings on the thermal bath; overall, these processes do not change the DM number density, but retain equilibrium among the dark species. 
Note that in \cref{eqs:BoltzmannEqs} we must use the thermally averaged rates, $\<\Gamma\>$. The thermal average introduces Lorentz dilation factors for decay processes -- which however are insignificant in the non-relativistic regime -- as well as Bose-enhancement factors in the case of transitions and capture or ionisation processes. We discuss this in more detail in \cref{sec:FreezeOut_CrossSection}. The thermally-averaged rates and cross-sections of inverse processes are related via detailed balance that we have already employed in writing \cref{eqs:BoltzmannEqs}, 
\begin{subequations}
\label{eq:DetailedBalance}
\label[pluralequation]{eqs:DetailedBalance}
\begin{align}
\<\Gamma_{\Bcal \to \Bcal'}^{\trans}\> &= 
\Gamma_{\Bcal'\to \Bcal}^{\trans} \times
(Y_{\Bcal'}^{\rm eq} / Y_{\Bcal}^{\rm eq}) ,
\label{eq:DetailedBalance_BB'}
\\
\<\Gamma_{\Bcal \to ij}^{\ion}\> &= s \,
\<\sigma_{ij \to \Bcal}^{\BSF} \vrel \> \times
(Y_i^{\rm eq}Y_j^{\rm eq} / Y_{\Bcal}^{\rm eq}) ,
\label{eq:DetailedBalance_IonBSF}
\\
\<\Gamma_{i\to j}\> &= 
\<\Gamma_{j\to i}\> \times (Y_j^{\rm eq} / Y_i^{\rm eq}) .
\label{eq:DetailedBalance_ij} 
\end{align}
\end{subequations}

The fractional relic DM density is
\begin{align}
\Omega \simeq (m-\sqrt{4\pi \aH}\vH) Y_\infty s_0 / \rho_c, 
\label{eq:OmegaDM}
\end{align} 
where $Y_\infty$ is the final yield, and 
we have included the mass shift of the lightest state that arises after the electroweak symmetry breaking (cf.~\cref{eq:MassEigenvalues}); this is significant only for the lower end of the mass range we consider and for large couplings $\aH$. In \cref{eq:OmegaDM}, $s_0 \simeq 2839.5~\cm^{-3}$ and $\rho_c \simeq 4.78 \cdot 10^{-6}~\GeV~\cm^{-3}$ are the entropy and critical energy densities of the universe today~\cite{Aghanim:2018eyx}.

\subsection{Effective Boltzmann equation \label{sec:FreezeOut_BoltzmannEff}}

The system of coupled Boltzmann \cref{eqs:BoltzmannEqs} is numerically difficult to solve. We shall thus adopt an effective method that reduces  \cref{eqs:BoltzmannEqs} to one equation for the DM yield \eqref{eq:DensityTotal_def}.

For convenience, we first define the total formation cross-section, ionisation rate and transition rate of every bound state ${\cal B}$,
\begin{subequations}
\label{eq:RatesTot_Def}
\label[pluralequation]{eqs:RatesTot_Def}
\begin{align}
\sigma_{\Bcal}^{\BSF}  &\equiv 
\sum_{i,j} \dfrac{g_i g_j}{g_{\DM}^2} 
\sigma_{ij \to \Bcal}^{\BSF} ,
\label{eq:sigma_BSF_tot_def}
\\
\Gamma_{\Bcal}^{\ion} &\equiv 
\sum_{i,j} \Gamma_{\Bcal \to ij}^{\ion},
\label{eq:Gamma_ion_tot_def}
\\
\Gamma_{\Bcal}^{\trans}  &\equiv 
\sum_{\Bcal'\neq \Bcal} \Gamma_{\Bcal \to \Bcal'}^{\trans}.
\label{eq:Gamma_trans_def}
\end{align}
\end{subequations}

We begin by assuming that the $i \leftrightarrow j$ interactions are sufficiently rapid to ensure chemical equilibrium among the free species, such that $Y_i / Y_i^{\eq} = w$, where $w$ is the same for all $i=S,D,\bar{D}$. Due to their rapid decays, inverse decays and transitions to other bound levels, the bound states are typically close to equilibrium, thus $dY_{\Bcal}/dx \simeq 0$. Under this assumption,  eqs.~\eqref{eq:BoltzmannEqs_Bound} yield a system of linear equations for $Y_{\Bcal}$ that can be solved and re-employed in \cref{eq:BoltzmannEqs_Free}~\cite{Ellis:2015vaa}. For bound states that do not participate in any bound-to-bound transitions, this simplifies to
\begin{align}
Y_{\Bcal} &= Y_{\Bcal}^{\eq} \ \dfrac
{\<\Gamma_{\Bcal}^{\dec}\> + w^2 \<\Gamma_{\Bcal}^{\ion}\> }
{\<\Gamma_{\Bcal}^{\dec}\> + \<\Gamma_{\Bcal}^{\ion}\> }.
\label{eq:YB_uncoupled}
\end{align}

In the model under consideration and within our approximations~\cite{Oncala:2021tkz},  the spin-1 $D\bar{D}$ and $DD$ bound states do not participate in any bound-to-bound transitions, while the spin-0 $SS/D\bar{D}$ and $DS$ bound states can rapidly transit into each other via $H$ absorption/emission (cf.~\cite[table~6]{Oncala:2021tkz}.) For the latter, eqs.~\eqref{eq:BoltzmannEqs_Bound} read
\begin{multline}
\left( 
\begin{array}{ccc}
\<\Gamma_{\SSDDbar}^{\dec}\> + \<\Gamma_{\SSDDbar}^\ion\> + 2\<\Gamma_{\SSDDbar \to \DS}^{\trans}\>
&~~~~~&
-2\<\Gamma_{\DS \to \SSDDbar}^{\trans}\>
\\
-\<\Gamma_{\SSDDbar \to \DS}^{\trans}\>
&~~~~~&
\<\Gamma_{\DS}^\ion\> + \<\Gamma_{\DS \to \SSDDbar}^{\trans}\>
\end{array}
\right)
\left(\begin{array}{c} Y_{\SSDDbar} \\ Y_{\DS} \end{array}\right) =
\\
= 
\left(
\begin{array}{c}
\[\<\Gamma_{\SSDDbar}^{\dec}\> + w^2 \<\Gamma_{\SSDDbar}^{\ion}\> \] Y_{\SSDDbar}^{\eq}
\\
w^2 \<\Gamma_{\DS}^{\ion}\> \, Y_{\DS}^{\eq}
\end{array}
\right),
\label{eq:YB_coupled}
\end{multline}
where we set $\<\Gamma_{\DS}^{\dec}\> \simeq 0$~\cite{Oncala:2021tkz}, 
and we recall that
$\<\Gamma_{\SSDDbar \to \DS}^{\trans}\> = 
\<\Gamma_{\DS \to \SSDDbar}^{\trans}\> 
(Y_{\DS}^{\eq} / Y_{\SSDDbar}^{\eq} )$,  
due to detailed balance \cref{eq:DetailedBalance_BB'}. The factors 2 in the first row account for transitions to and from the two conjugate bound states $DS$ and $\bar{D}S$. 

Next, we use \cref{eq:YB_uncoupled} for the $D\bar{D}$ and $DD$ yields and the solution of \cref{eq:YB_coupled} for the $SS/D\bar{D}$ and $DS$ yields, in the Boltzmann \cref{eq:BoltzmannEqs_Free}. Summing over all free particle species, we find that the evolution of $Y$ is governed by the Boltzmann equation
\begin{align}
\frac{dY}{dx} = -\sqrt{\frac{\pi}{45}} 
\, \dfrac{\mpl \, m \, g_{*,\rm eff}^{1/2}}{x^2}
\, \<\sigma \vrel\>_\eff  
\: [Y^2 - (Y^{\eq})^2] ,
\label{eq:BoltzmannEq_Eff}
\end{align}
where the equilibrium density $Y^{\eq}$ is given in \cref{eq:Yeq}. The DM destruction cross-section $\<\sigma \vrel\>_\eff$ receives contributions from direct annihilation and BSF processes, 
\begin{align}
\< \sigma \vrel \>_\eff = 
\< \sigma_\ann \vrel \> +  \< \sigma_\BSF \vrel \>_\eff ,
\label{eq:FreezeOut_sigmavEff_def}
\end{align}
with
\begin{align}
\< \sigma_\ann \vrel \> \equiv \sum_{i,j} 
\dfrac{g_i g_j}{g_{\DM}^2} \< \sigma_{ij}^\ann \vrel \> ,
\label{eq:FreezeOut_sigmavAnn_def}
\end{align}
and 
\begin{subequations}
\label{eq:FreezeOut_sigmavBSFeff_def}
\label[pluralequation]{eqs:FreezeOut_sigmavBSFeff_def}
\begin{align}
\< \sigma_\BSF \vrel \>_\eff  
= \<\sigma_{\SSDDbar}^{\BSF} \vrel \>_\eff  
+ \<\sigma_{\DDbar}^{\BSF} \vrel \>_\eff  
+ 2\<\sigma_{\DD}^{\BSF} \vrel \>_\eff  
+ 2\<\sigma_{\DS}^{\BSF} \vrel \>_\eff  , 
\tag{\ref{eq:FreezeOut_sigmavBSFeff_def}}
\label{eq:FreezeOut_sigmavBSFeff}
\end{align}
where the factors 2 in the $DD$ and $DS$ terms account also for the formation of the conjugate bound states. The individual contributions are found as follows. For the bound-states that do not participate in any bound-to-bound transitions,
\begin{align}
\dfrac
{\<\sigma_{\DDbar}^{\BSF} \vrel \>_\eff}
{\<\sigma_{\DDbar}^{\BSF} \vrel \>}  
&=
\dfrac{\<\Gamma_{\DDbar}^{\dec}\> }{ \<\Gamma_{\DDbar}^{\dec}\> + \<\Gamma_{\DDbar}^{\ion}\> } ,
\label{eq:FreezeOut_sigmavBSFeff_DDbar} 
\\
\dfrac
{\<\sigma_{\DD}^{\BSF} \vrel \>_\eff}
{\<\sigma_{\DD}^{\BSF} \vrel \>} 
&=
\dfrac{\<\Gamma_{\DD}^{\dec}\> }{ \<\Gamma_{\DD}^{\dec}\> + \<\Gamma_{\DD}^{\ion}\> } ,
\label{eq:FreezeOut_sigmavBSFeff_DD} 
\end{align}
while for the coupled bound states
\begin{align}
\dfrac
{\<\sigma_{\SSDDbar}^{\BSF} \vrel \>_\eff}
{\<\sigma_{\SSDDbar}^{\BSF} \vrel \>}  
&=
\dfrac{\<\Gamma_{\SSDDbar}^{\dec}\> }
{ 
\<\Gamma_{\SSDDbar}^{\dec}\> + \<\Gamma_{\SSDDbar}^{\ion}\> 
+2\dfrac{ \<\Gamma_{\DS}^{\ion}\> \, \<\Gamma_{\DS\to\SSDDbar}^{\trans}\> }
{ \<\Gamma_{\DS}^{\ion}\> + \<\Gamma_{\DS\to\SSDDbar}^{\trans}\> }
\dfrac{Y_{\DS}^{\eq}}{Y_{\SSDDbar}^{\eq}}
} ,
\label{eq:FreezeOut_sigmavBSFeff_SSDDbar} 
\\
\dfrac
{\<\sigma_{\DS}^{\BSF} \vrel \>_\eff}
{\<\sigma_{\DS}^{\BSF} \vrel \>} 
&=
\dfrac
{\<\Gamma_{\DS\to\SSDDbar}^{\trans}\> \times
	\dfrac{\<\Gamma_{\SSDDbar}^{\dec}\>}{\<\Gamma_{\SSDDbar}^{\dec}\> + \<\Gamma_{\SSDDbar}^{\ion}\>} }
{ 
\<\Gamma_{\DS}^{\ion}\> + \<\Gamma_{\DS\to \SSDDbar}^{\trans}\>
+2 \dfrac{ \<\Gamma_{\DS}^{\ion}\> \<\Gamma_{\DS\to\SSDDbar}^{\trans}\> } 
{\<\Gamma_{\SSDDbar}^{\dec}\> + \<\Gamma_{\SSDDbar}^{\ion}\> }
\ \dfrac{Y_{\DS}^{\eq}}{Y_{\SSDDbar}^{\eq}}
} .
\label{eq:FreezeOut_sigmavBSFeff_DS} 
\end{align}
\end{subequations}
In \cref{eqs:FreezeOut_sigmavBSFeff_def}, $\<\sigma_{\Bcal}^{\BSF}\vrel\>$ are the thermal averages of the actual velocity-weighted formation cross-sections for every bound state, defined in \cref{eq:sigma_BSF_tot_def}; we discuss them further in the following section. Note that if the transitions between the $SS/D\bar{D}$ and $DS$ bound states are very rapid, in particular when $\<\Gamma_{\DS\to\SSDDbar}^{\trans}\> \gg \<\Gamma_{\DS}^{\ion}\>$, then the branching ratios that weigh their actual BSF cross-sections in \cref{eq:FreezeOut_sigmavBSFeff_SSDDbar,eq:FreezeOut_sigmavBSFeff_DS}  are equal.

\subsection{Effective cross-section \label{sec:FreezeOut_CrossSection}}

\begin{figure}[t!]
\centering
\includegraphics[width=0.98\textwidth]{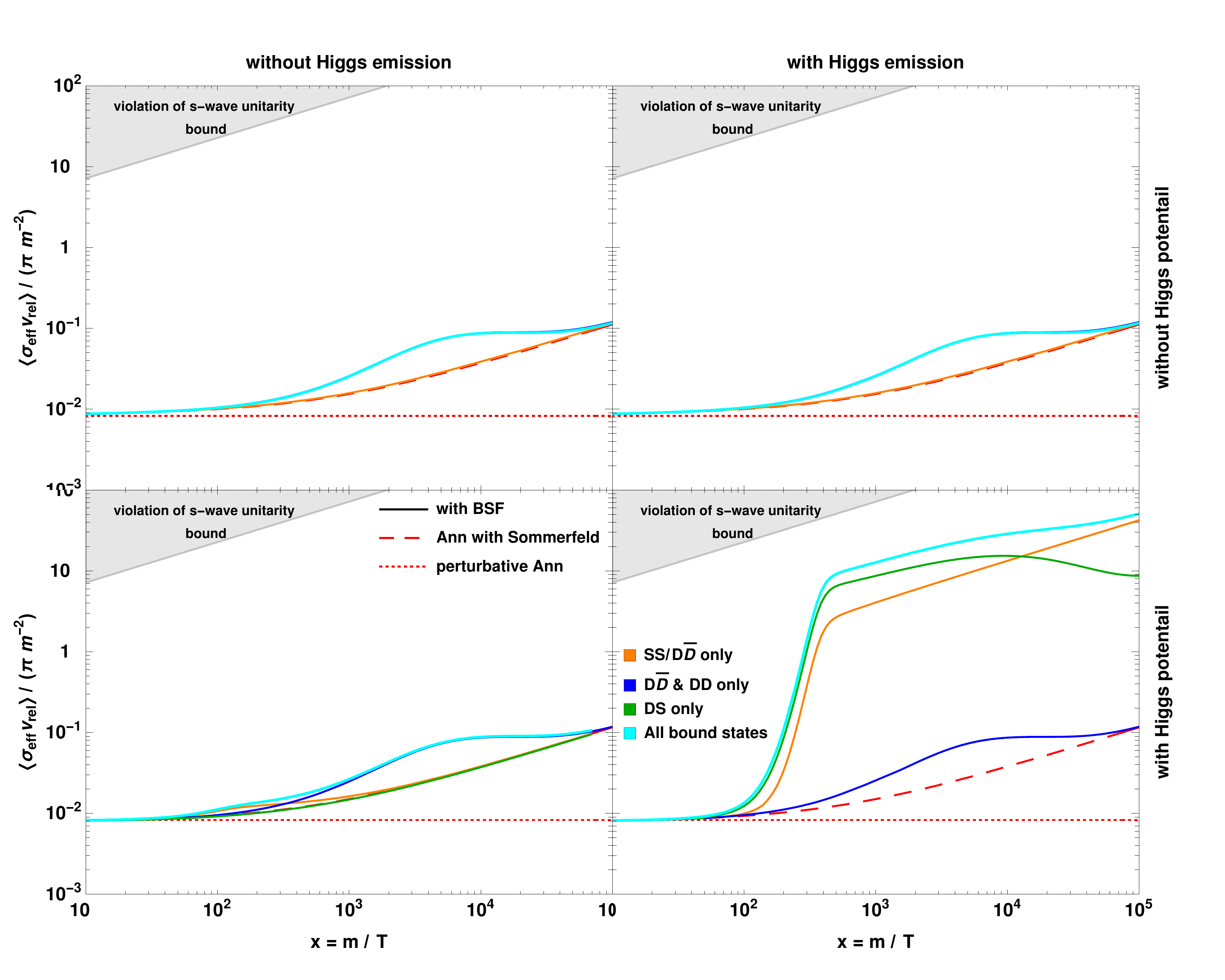}
\caption{\label{fig:EffectiveCrossSection_Contributions} 
Contributions to the effective DM destruction cross-section. The solid lines include direct annihilation with Sommerfeld effect plus BSF according to the colour legend. We have used $m=20~\TeV$, $\aH=0.2$ and the temperature dependent Higgs mass $\mH(T)$. Note that the cross-sections have been regulated according to \cite[section~3.6.]{Oncala:2021tkz}.   
The binding energy of the $D\bar{D}$ and $DD$ bound states does not depend on the coupling to the Higgs, and their formation via $H$ emission or exchange is always suppressed due the Higgs mass; their contribution is dominated by $W$ emission. Both the $SS/D\bar{D}$ and $DS$ binding energies depend on $\aH$, which ensures that their formation via $H$ emission is not suppressed when the Higgs-mediated potential is taken into account and provided that $\aH$ is sufficiently large (bottom right panel.) The $DS$ bound states do not exist when neglecting the Higgs-mediated potential (upper row.)}
\end{figure}

We now consider in more detail the contributions to the effective DM destruction cross-section in our model, based on the computations of ref.~\cite{Oncala:2021tkz}. 
We begin with direct annihilation in \cref{sec:FreezeOut_CrossSection_Annihilation}, and then discuss BSF in \cref{sec:FreezeOut_CrossSection_BSF}. In \cref{fig:BSF_CrossSections_ThermalAverage_DDbarDD,fig:BSF_CrossSections_ThermalAverage_SSDDbarDS} we illustrate the contributions to BSF, while in \cref{fig:EffectiveCrossSection_Contributions} we compare all contributions to the DM destruction cross-section for a chosen set of parameters, showcasing the effect of the Higgs potential and of BSF via Higgs emission.

\subsubsection{Annihilation \label{sec:FreezeOut_CrossSection_Annihilation}}

In our model, the total annihilation cross-section is
\begin{align}
\sigma_\ann \vrel 
&= \[
   g_{\SS}		(\sigma^\ann_{\SS} \vrel)  
+ 2g_{\DDbar}	(\sigma^\ann_{\DDbar} \vrel) 
+ 2g_{\DD}		(\sigma^\ann_{\DD} \vrel)  
+ 4g_{\DS}		(\sigma^\ann_{\DS} \vrel)  
\]/ g_{\DM}^2,
\label{eq:FreezeOut_sigmav_Ann_Total}  
\end{align}
where the indices denote the two-particle scattering states, with dof 
$g_{\SS} = 4$, 	
$g_{\DD} = 16$, 
$g_{\DDbar} = 16$, 
$g_{\DS} = 8$. 
The $DD$ and $DS$ contributions carry factors of 2 to account also for the annihilation of the conjugate states, and $D\bar{D}$ and $DS$ carry factors of 2 to account for the two distinguishable particles annihilated in each process.\footnote{As is well known, for pairs of identical particles (here $SS$, $DD$, $\bar{D}\bar{D}$), this factor is canceled upon thermal averaging by the factor 1/2 needed to avoid double-counting of the initial particle states~\cite{Gondolo:1990dk}.} 
From \cite[table~5]{Oncala:2021tkz}, we find
\begin{subequations}
\label{eq:FreezeOut_sigmav_Ann}
\label[pluralequation]{eqs:FreezeOut_sigmav_Ann}
\begin{align}
g_{\SS}
(\sigma^\ann_{\SS} \vrel) /(\pi m^{-2}) &= 0 ,
\label{eq:FreezeOut_sigmav_Ann_SS}
\\
g_{\DDbar}
(\sigma^\ann_{\DDbar} \vrel) /(\pi m^{-2}) 
&=
1
\times \(\frac{\alpha_1^2}{2}+\frac{3\alpha_2^2}{2}\) 
\times \frac{\aA S_0(\zetaA)+\aR S_0(-\zetaR)}{\aA+\aR}
\nn \\
&+
3
\times \[ \frac{(\alpha_1+2\aH)^2}{12} + \frac{10\alpha_1^2}{3} \] 
\times S_0 \(\frac{\zeta_1+3\zeta_2}{4}\)
\nn \\
&+
3
\times \alpha_1 \alpha_2  
\times S_0 \(\frac{\zeta_1-\zeta_2}{4}\)
\nn \\
&+
9
\times \[ \frac{(\alpha_2+2\aH)^2}{12} + \frac{\alpha_2^2}{12} + 2 \alpha_2^2 \]
\times S_0 \(\frac{\zeta_1-\zeta_2}{4}\) ,
\label{eq:FreezeOut_sigmav_Ann_DDbar}
\\
g_{\DD}
(\sigma^\ann_{\DD} \vrel) /(\pi m^{-2}) &=
3
\times \frac{4\aH^2}{3} 
\times S_0\( \frac{-\zeta_1+3\zeta_2}{4} \) ,
\label{eq:FreezeOut_sigmav_Ann_DD}
\\
g_{\DS} 
(\sigma^\ann_{\DS} \vrel) /(\pi m^{-2}) 
&=
6 
\times \(\frac{\alpha_1 \aH}{6} +\frac{\alpha_2 \aH}{2}\) 
\times S_0 (-\zetaH) .
\label{eq:FreezeOut_sigmav_Ann_DS}
\end{align}
\end{subequations}
In the above, the $s$-wave Sommerfeld factor is
\begin{align}
S_0 (\zetaS) \equiv \dfrac{2\pi \zetaS}{1-e^{-2\pi \zetaS}} ,
\label{eq:S0_def}
\end{align}
where $\zetaS \equiv \aS /\vrel$, with $\aS$ being the strength of the Coulomb potential of the scattering state. The various $\zetaS$ appearing in \cref{eqs:FreezeOut_sigmav_Ann} are
\begin{align}
\zeta_1 \equiv \alpha_1/\vrel, \quad
\zeta_2 \equiv \alpha_2/\vrel, \quad
\zetaH \equiv \aH/\vrel, \quad
\zetaA \equiv \aA/\vrel, \quad
\zetaR \equiv \aR/\vrel.
\label{eq:zetas_def}
\end{align}

Each of the terms in \cref{eqs:FreezeOut_sigmav_Ann} is the product of the dof, the perturbative annihilation cross-section and the Sommerfeld factor of a spin- and gauge-projected state. The $D\bar{D}$ cross-section includes contributions from both the $\SUL$ singlet and triplet projections, which are characterised by different non-relativistic potentials and thus have different Sommerfeld factors. For the singlet states, the potential depends also on the spin. Indeed, the spin-0 $\SUL$ singlet $SS$ and $D\bar{D}$ states mix due to the Higgs mediated potential. Since the perturbative $s$-wave annihilation of the $SS$ component vanishes, the contribution from the annihilation of the $SS$-like state has been included in the $D\bar{D}$-like state, for simplicity. We refer to \cite{Oncala:2021tkz} for more details.

The thermally averaged annihilation cross-section \eqref{eq:FreezeOut_sigmavAnn_def} is found from \eqref{eq:FreezeOut_sigmav_Ann_Total}, \eqref{eq:FreezeOut_sigmav_Ann} and
\begin{align}
\<\sigma_\ann \vrel \> = \(\frac{m}{4\pi T}\)^{3/2}
\int d^3 \vrel \, e^{-m \vrel^2/(4T)} 
\ (\sigma_\ann \vrel) .
\label{eq:FreezeOut_sigmav_Ann_Averaged}  
\end{align}

\subsubsection{Bound state formation \label{sec:FreezeOut_CrossSection_BSF}}

The radiative BSF cross-sections have been summarised in \cite[tables~7--10]{Oncala:2021tkz}, and we shall denote them here as $\sigma_{\Bcal}^{\rBSF}\vrel [xx]$ with $xx$ and ${\cal B}$ being the scattering and bound states. Bound states can also form through scattering on the thermal bath, via exchange of an off-shell mediator; the corresponding cross-section factorise in their radiative counterparts and temperature-dependent functions~\cite{Binder:2019erp,Binder:2020efn,Oncala:2021tkz}. Collecting these results, the velocity-weighted cross-sections $\sigma_{\Bcal}^{\BSF} \vrel$ for the formation of the various bound-state species ${\cal B}$ receive the following contributions from the individual channels (cf.~\cite[tables~7-10]{Oncala:2021tkz})
\begin{subequations}
\label{eq:FreezeOut_sigmav_BSF}
\label[pluralequation]{eqs:FreezeOut_sigmav_BSF}
\begin{align}
\sigma_{\SSDDbar}^{\BSF} \vrel  = \frac{1}{g_{\DM}^2} 
&\left\{
2 \times 1 \times (1+\RB) \times
\sigma_{\SSDDbar}^{\rBSF}\vrel   
[(D\bar{D})_{(\mathbb{1},0)}^{\text{spin-0}}]
\right. \nn \\ &\left.
+2 \times 3 \times (1+\RW) \times
\sigma_{\SSDDbar}^{\rBSF}\vrel   
[(D\bar{D})_{(\mathbb{3},0)}^{\text{spin-0}}]
\right. \nn \\ &\left.
+4 \times 2 \times (1+\RH/\hH) \times
\sigma_{\SSDDbar}^{\rBSF}\vrel   
[(DS)_{(\mathbb{2},1/2)}^{\text{spin-0}} ] 
\right\} ,
\label{eq:FreezeOut_sigmav_BSF_SSDDbar}
\end{align}
\begin{align}
\sigma_{\DDbar}^{\BSF} \vrel  =  \frac{1}{g_{\DM}^2} 
&\left\{
 2 \times 3 \times (1+\RB) \times 
\sigma_{\DDbar}^{\rBSF}\vrel   [(D\bar{D}\text{-like})_{(\mathbb{1},0)}^{\text{spin-1}}]
\right. \nn \\ &\left.
+1 \times 3 \times (1+\RB) \times
\sigma_{\DDbar}^{\rBSF}\vrel   [(SS\text{-like})_{(\mathbb{1},0)}^{\text{spin-1}}]
\right. \nn \\ &\left.
+2 \times 9 \times (1+\RW) \times 
\sigma_{\DDbar}^{\rBSF}\vrel   [(D\bar{D})_{(\mathbb{3},0)}^{\text{spin-1}}]
\right. \nn \\ &\left.
+4 \times 6 \times (1+\RH/\hH) \times
\sigma_{\DDbar}^{\rBSF}\vrel   [(DS)_{(\mathbb{2},1/2)}^{\text{spin-1}}] 
\right\} ,
\label{eq:FreezeOut_sigmav_BSF_DDbar}
\end{align}
\begin{align}
\sigma_{\DD}^{\BSF} \vrel  =  \frac{1}{g_{\DM}^2} 
&\left\{
1 \times 9 \times (1+\RW) \times  
\sigma_{\DD}^{\rBSF}\vrel   
[(DD)_{(\mathbb{3},1)}^{\text{spin-1}}]
\right. \nn \\ &\left.
+2 \times 6 \times (1+\RH/\hH) \times
\sigma_{\DD}^{\rBSF}\vrel   
[(DS)_{(\mathbb{2},1/2)}^{\text{spin-1}}] 
\right\},
\label{eq:FreezeOut_sigmav_BSF_DD}
\end{align}
\begin{align}
\sigma_{\DS}^{\BSF} \vrel  =  \frac{1}{g_{\DM}^2} 
&\left\{
2 \times 2 \times (1+\RB) \times  
\sigma_{\DS}^{\rBSF}\vrel   [(DS)_{(\mathbb{2},1/2)}^{\text{spin-0}},~~B~\text{emission} ]
\right. 
\nn \\ &
+2 \times 2 \times (1+\RW) \times  
\sigma_{\DS}^{\rBSF}\vrel   [(DS)_{(\mathbb{2},1/2)}^{\text{spin-0}},~~W~\text{emission} ]
\nn \\ &
+1 \times 1 \times (1+\RH/\hH) \times
\sigma_{\DS}^{\rBSF}\vrel   [(SS\text{-like})_{(\mathbb{1},0)}^{\text{spin-0}} ]
\nn \\ &
+2 \times 1 \times (1+\RH/\hH) \times
\sigma_{\DS}^{\rBSF}\vrel   [(D\bar{D}\text{-like})_{(\mathbb{1},0)}^{\text{spin-0}} ]
\nn \\ &
+2 \times 3 \times (1+\RH/\hH) \times
\sigma_{\DS}^{\rBSF}\vrel   [(D\bar{D})_{(\mathbb{3},0)}^{\text{spin-0}} ]
\nn \\ &
\left.
+1 \times 3 \times (1+\RH/\hH) \times
\sigma_{\DS}^{\rBSF}\vrel   [(DD)_{(\mathbb{3},1)}^{\text{spin-0}} ] 
\right\}.
\label{eq:FreezeOut_sigmav_BSF_DS}
\end{align}
\end{subequations}
In each term above, the first factor accounts for the number of DM particles destroyed (upon thermal averaging), as well as the capture of the conjugate scattering state if applicable, in analogy to \cref{eq:FreezeOut_sigmav_Ann_Total} for annihilation. The second factor corresponds to the dof of the scattering state. 

The factors in the brackets sum the radiative and via-scattering contributions to BSF. The factors $\RH$, $\RB$, $\RW$ indicate the off-shell exchange of $H$, $B$ and $W$ bosons with the thermal bath, and depend on $\omega/T$, where 
\begin{align}
\omega = \mu (\aB^2 + \vrel^2)/2
\label{eq:omega}
\end{align} 
is the energy dissipated in the capture process~\cite{Petraki:2015hla,Oncala:2021tkz}, with $\aB$ being the strength of the potential of the corresponding bound state (cf.~\cref{tab:BoundStatesGroundLevel} and \cite[table~6]{Oncala:2021tkz}.) $\RH$ depends also on $\mH/\omega$, and essentially replaces the phase-space suppression 
\begin{align}
\hH (\omega) \equiv \(1-\mH^2/\omega^2\)^{1/2}
\label{eq:hH}
\end{align} 
due to on-shell $H^{(\dagger)}$ emission that is included in the radiative cross-sections. 
The $\RH$, $\RB$, $\RW$ factors can be found in \cite{Oncala:2021tkz}.

Next, we must thermally average \cref{eqs:FreezeOut_sigmav_BSF}. In BSF, the emitted boson carries away a small amount of energy that can be comparable to the temperature of the primordial plasma during the DM decoupling. The Bose enhancement due to the final state boson can thus be significant, and must be included in thermal averaging the BSF cross-sections to ensure that detailed balance holds~\cite{vonHarling:2014kha},\footnote{We recall from \cite{Oncala:2021tkz} that a factor of $[1+1/(e^{\omega/T}-1)]$ has been pulled out from the definition of the BSF cross-section via scattering, such that \cref{eq:FreezeOut_sigmav_BSF_Averaged} is the appropriate thermal-averaging formula for both radiative BSF and BSF via scattering.}  
\begin{align}
\<\sigma_{\Bcal}^{\BSF} \vrel \> = \(\frac{m}{4\pi T}\)^{3/2}
\int d^3 \vrel \, e^{-m \vrel^2/(4T)} 
\(1 + \frac{1}{e^{\omega/T}-1}\)
\ (\sigma_{\Bcal}^{\BSF} \vrel) ,
\label{eq:FreezeOut_sigmav_BSF_Averaged}  
\end{align}

\medskip

As seen in \cref{eq:FreezeOut_sigmavBSFeff_def}, the contributions of each bound level to the effective DM destruction cross-section \eqref{eq:FreezeOut_sigmavEff_def} have to be waited by the appropriate branching fractions that account for the portion of bound states that decay into radiation thereby reducing the DM density. The bound-state decay and transition rates needed to compute these branching fractions can be found in \cite[table~6]{Oncala:2021tkz}. In thermally averaging these rates, we may neglect the Lorentz dilation factor that is $\simeq 1$ in the non-relativistic regime. However, the low-energy boson emitted in bound-to-bound transitions implies a Bose enhancement that must be included to ensure detailed balance at temperatures higher than the dissipated energy. So, 
\begin{subequations}
\label{eq:ThermalAverage_Rates}
\label[pluralequation]{eqs:ThermalAverage_Rates}
\begin{align}
\<\Gamma_{\Bcal}^{\dec} \> &\simeq \Gamma_{\Bcal}^{\dec} ,
\label{eq:ThermalAverage_Decay} 
\\
\<\Gamma_{\DS \to \SSDDbar}^{\trans} \> &\simeq 
[1+ \RH (\omega)/ \hH (\omega) ] 
\(1+\dfrac{1}{e^{\omega/T}-1}\)
\, \Gamma_{\DS \to \SSDDbar}^{\trans}  ,
\label{eq:ThermalAverage_Transition}
\end{align}
where the dissipated energy here is $\omega =m(\aA^2-\aH^2)/4$, and $\hH(\omega)$ is the phase-space suppression defined in \cref{eq:hH}. Finally, the bound-state ionisation rates are computed using the detailed balance \cref{eq:DetailedBalance_IonBSF}, and summing over all ionised states as in \cref{eq:sigma_BSF_tot_def,eq:Gamma_ion_tot_def}; this yields
\begin{align}
\<\Gamma_{\Bcal}^{\ion}\> \simeq  
\<\sigma_{\Bcal}^\BSF \vrel\> \times 
\frac{g_{\DM}^2}{g_{\Bcal}}
\, \(\dfrac{mT}{4\pi}\)^{3/2}
\, e^{-|{\cal E}_{\Bcal}|/T}. 
\label{eq:ThermalAverage_Ion}
\end{align}
\end{subequations}

\subsubsection*{Ionisation equilibrium}

\Cref{eq:ThermalAverage_Ion} implies that at $T \gg |{\cal E}_{\Bcal}|$, the ionisation of the bound states tends to be faster than their decays and transitions, i.e.~$\<\Gamma_{\Bcal}^{\ion}\> \gg 
\<\Gamma_{\Bcal}^{\dec}\>,  \<\Gamma_{\Bcal}^{\trans}\>$, provided that  $\<\sigma_{\Bcal}^\BSF \vrel\>$ is sufficiently large. If so, the system reaches a state of \emph{ionisation equilibrium}, where the effective BSF cross-sections \eqref{eq:FreezeOut_sigmavBSFeff_def} become independent of the actual ones~\cite{Binder:2018znk}, 
\begin{align}
\<\sigma_{\Bcal}^{\BSF} \vrel\>_{\eff} \simeq
\dfrac{g_{\Bcal}^{}}{g_{\DM}^2} 
\, \Gamma^{\dec}_{\Bcal} \(\dfrac{4\pi}{mT}\)^{3/2}
e^{+|{\cal E}_{\Bcal}| / T} ,
\label{eq:IonisationEquilibrium}
\end{align}
where for the $DS$ bound state whose direct decay into radiation is suppressed, we must use the effective decay rate (cf.~\cref{eq:FreezeOut_sigmavBSFeff_DS})
\begin{align}
\<\Gamma_{\DS}^{\dec}\> \to 
\<\Gamma_{\DS\to\SSDDbar}^{\trans}\>
\dfrac{\<\Gamma_{\SSDDbar}^{\dec}\>}
{\<\Gamma_{\SSDDbar}^{\ion}\> + \<\Gamma_{\SSDDbar}^{\dec}\>
+2 \<\Gamma_{\DS\to \SSDDbar}^{\trans} \> 
(Y_{\DS}^{\eq} / Y_{\SSDDbar}^{\eq})} .
\label{eq:DecayRate_DS_effective}
\end{align}
Since the bound-state decay rates are proportional to the annihilation cross-sections of the corresponding scattering states (cf.~e.g.~ref.~\cite{Oncala:2021tkz}), \cref{eq:IonisationEquilibrium} implies that at high temperatures and while ionisation equilibrium holds, the BSF contribution to the DM destruction rate is negligible in comparison to that of direct annihilation (cf.~e.g.~\cite[eq.~(3.20)]{Oncala:2019yvj}.)

However, as $T$ approaches or drops below $|{\cal E}_{\Bcal}|$, the ionisation rates become exponentially suppressed and are overcome by the bound-state decay and/or bound-to-bound transition rates. For the uncoupled bound states $D\bar{D}$ and $DD$, this implies that the effective BSF cross-sections increase exponentially until they saturate to their maximum values, the actual BSF cross-sections.  
For the $SS/D\bar{D}$ and $DS$ coupled system, 
$\<\Gamma_{\DS\to \SSDDbar}^{\trans} \> > \<\Gamma_{\DS}^{\ion} \>$, occurs before the decay rates surpass the ionisation rates; in this interval, the effective BSF cross-sections \eqref{eqs:FreezeOut_sigmavBSFeff_def} together with the detailed balance \cref{eq:ThermalAverage_Ion}, imply that ionisation equilibrium holds for the sum of the $SS/D\bar{D}$ and $DS$ contributions,
\begin{align}
\<\sigma_{\SSDDbar}^{\BSF} \vrel \>_{\eff} +
2\<\sigma_{\DS}^{\BSF} \vrel \>_{\eff} \simeq 
\dfrac{g_{\SSDDbar}^{}}{g_{\DM}^2} 
\, \Gamma^{\dec}_{\SSDDbar} \(\dfrac{4\pi}{mT}\)^{3/2}
e^{+|{\cal E}_{\SSDDbar}| / T} ,
\label{eq:IonisationEquilibrium_SSDDbarAndDS_Couupled}
\end{align}
where again we neglected the $SS/D\bar{D}$ decay against ionisation rate. 
At even lower temperatures, when ionisation becomes slower than decay, the effective BSF cross-sections reach their actual values.

\begin{figure}[t!]
\centering
\includegraphics[width=\textwidth]{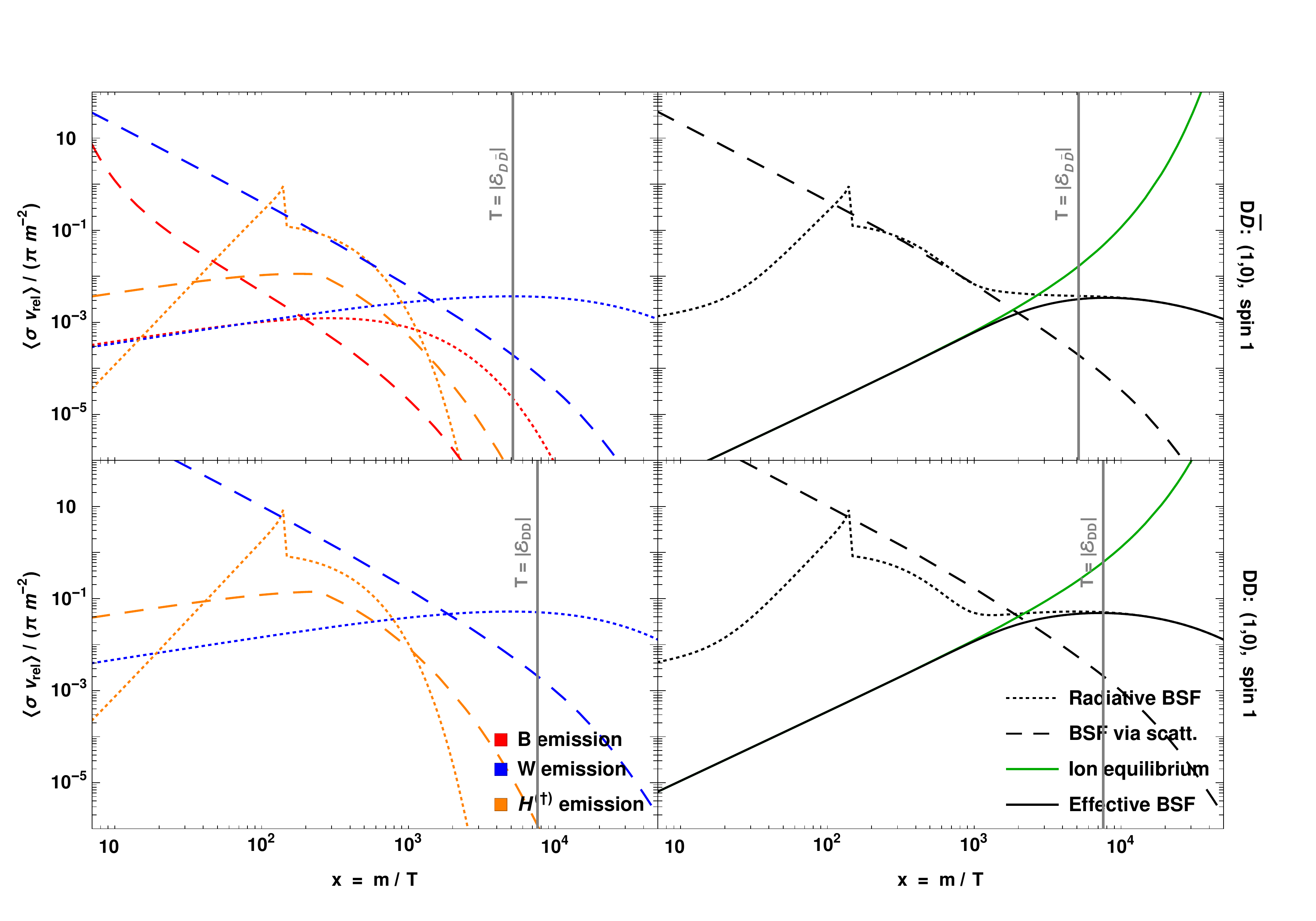}
\caption{\label{fig:BSF_CrossSections_ThermalAverage_DDbarDD} 
Thermally averaged BSF cross-sections for the $D\bar{D}$ and $DD$ bound states; the latter includes the capture into its conjugate. We have used $m=20~\TeV$, $\aH=0.2$ and the temperature dependent Higgs mass $\mH(T)$; the spikes in the radiative BSF occur at the EWPT, when the Higgs mass tends to zero before becoming $\mh \simeq 125~\GeV$. The vertical lines mark the temperatures equal to the binding energies. Note that the cross-sections have been regulated according to \cite[section~3.6.]{Oncala:2021tkz}.}
\end{figure}
\begin{figure}[t!]
\centering
\includegraphics[width=\textwidth]{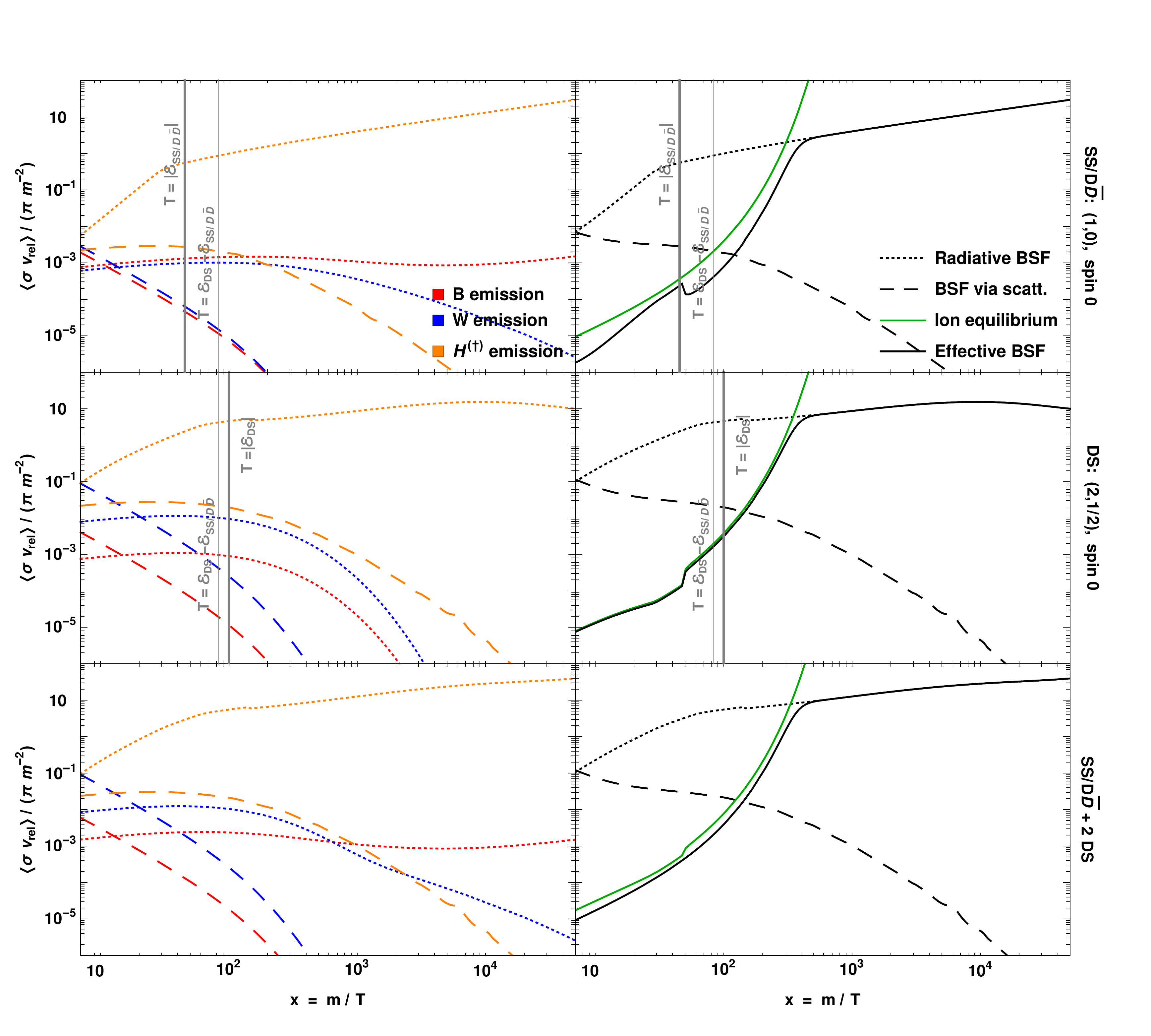}
\caption{\label{fig:BSF_CrossSections_ThermalAverage_SSDDbarDS} 
As in \cref{fig:BSF_CrossSections_ThermalAverage_DDbarDD}, but for the $SS/D\bar{D}$ and $DS$ states that transition into each other via Higgs emission or absorption. The $DS$ panels include the capture into its conjugate. In the bottom row, we show the sum of the $SS/D\bar{D}$, $DS$ and $\bar{D}S$ contributions. The vertical lines mark the temperatures equal to the binding energies and the energy splitting between $SS/D\bar{D}$ and $DS$. The feature around $x \simeq 50$ occurs when the Higgs doublet mass becomes lower than the energy splitting between the two bound states; this opens up the bound-to-bound transitions via on-shell Higgs emission (at higher $T$ they occur only via off-shell Higgs exchange with the thermal bath), and drives the $SS/D\bar{D}$ bound states somewhat away from ionisation equilibrium.  
}
\end{figure}

We illustrate the above in \cref{fig:BSF_CrossSections_ThermalAverage_DDbarDD,fig:BSF_CrossSections_ThermalAverage_SSDDbarDS}, where we also compare radiative BSF and BSF via scattering. Two observations are useful more generally for calculations of freeze-out with bound states:
\begin{itemize}
\item	
In some (but not all) cases, BSF via scattering dominates at early times; BSF via Higgs exchange may also dominate at late-times over on-shell emission due to the phase-space suppression of the latter. Nevertheless BSF via scattering does not change significantly the effective BSF cross-section with respect to considering radiative BSF only, because overall it becomes subdominant  while the system is still in ionisation equilibrium, or around the time it exits it.

\item 
For the $D\bar{D}$, $DD$ bound states, ionisation equilibrium ceases at $T > |{\cal E}_{\Bcal}|$ (cf. \cref{fig:BSF_CrossSections_ThermalAverage_DDbarDD}.) 

In contrast, the bound-to-bound transitions prevent the $SS/D\bar{D}$ and $DS$ coupled system to reach ionisation equilibrium. However, it closely tracks it until much lower temperatures, $T \ll |{\cal E}_{\Bcal}|$, due to the largeness of the BSF cross-sections  (cf.~\cref{fig:BSF_CrossSections_ThermalAverage_SSDDbarDS}.) 

We also note here that the computation of the DM thermal decoupling (cf.~\cref{Sec:FreezeOut_Results}) shows that much of the BSF effect on the relic density arises after the system exits ionisation equilibrium. (This was also found in ref.~\cite{vonHarling:2014kha}.) 

The above imply that it is \emph{not} safe to estimate the BSF effect by assuming ionisation equilibrium until $T \sim |{\cal E}_{\Bcal}|$ and neglecting any effect thereafter, an approach previously adopted in refs.~\cite{Kim:2016zyy,Biondini:2017ufr,Biondini:2018pwp,Biondini:2018ovz,Biondini:2019int}. Considering instead the BSF cross-sections is necessary for an accurate computation. 

\end{itemize}

\clearpage
\section{Results \label{Sec:FreezeOut_Results}}

\subsection{Timeline and relic density \label{sec:FreezeOut_RelicDensity}}

Collecting all the above, we are now ready to compute the DM decoupling and relic density. We consider and compare the cases described in \cref{tab:CasesForComparison}, and recall that our calculations always assume electroweak symmetry. We discuss this approximation in \cref{sec:FreezeOut_Approximations}.

\begin{table}[h!]
\renewcommand*{\arraystretch}{1.8}	
\begin{tabular}{|l|l|}
\hline 	
{\bf AnnS${}_{\BW}$} 
& 
\parbox{8.75cm}{Annihilation with Sommerfeld effect \\ 
due to the $B,W$-mediated potentials.}
\\[1.5ex] \hline
{\bf AnnS${}_{\BW}$ + $\BW$-BSF${}_{\BW}$}
& 
\parbox{8.75cm}{Annihilation and BSF via $B$ or $W$ emission or\\ 
exchange, including the $B,W$-mediated potentials.}
\\[1.5ex] \hline 
{\bf AnnS${}_{\BWH}$ + $\BWH$-BSF${}_{\BWH}$}
& 
\parbox{8.75cm}{Annihilation and BSF via on-shell $B$, $W$ or $H^{(\dagger)}$ \\ 
emission, including the $B,W,H$-mediated potentials.}
\\[1.5ex] \hline 
\end{tabular}
\\[1em]
\caption{\label{tab:CasesForComparison}
The combinations of effects we compare in the following, in terms of their impact on the DM decoupling. The case of perturbative annihilation only does not differ very significantly from annihilation with Sommerfeld effect, thus we do not present it separately. When considering BSF, we always include both radiative BSF and BSF via scattering. However, we have examined their effects separately, and found that the inclusion of BSF via scattering does not change the results obtained when considering radiative BSF only.  Moreover, in the present model, considering the Higgs-mediated potential while omitting BSF via Higgs emission, or the reverse, do not result in a significant effect on the relic density (cf.~\cref{fig:EffectiveCrossSection_Contributions}), we thus do not present these cases separately.}
\end{table}

\begin{figure}[t!]
\centering
\includegraphics[width=0.98\textwidth]{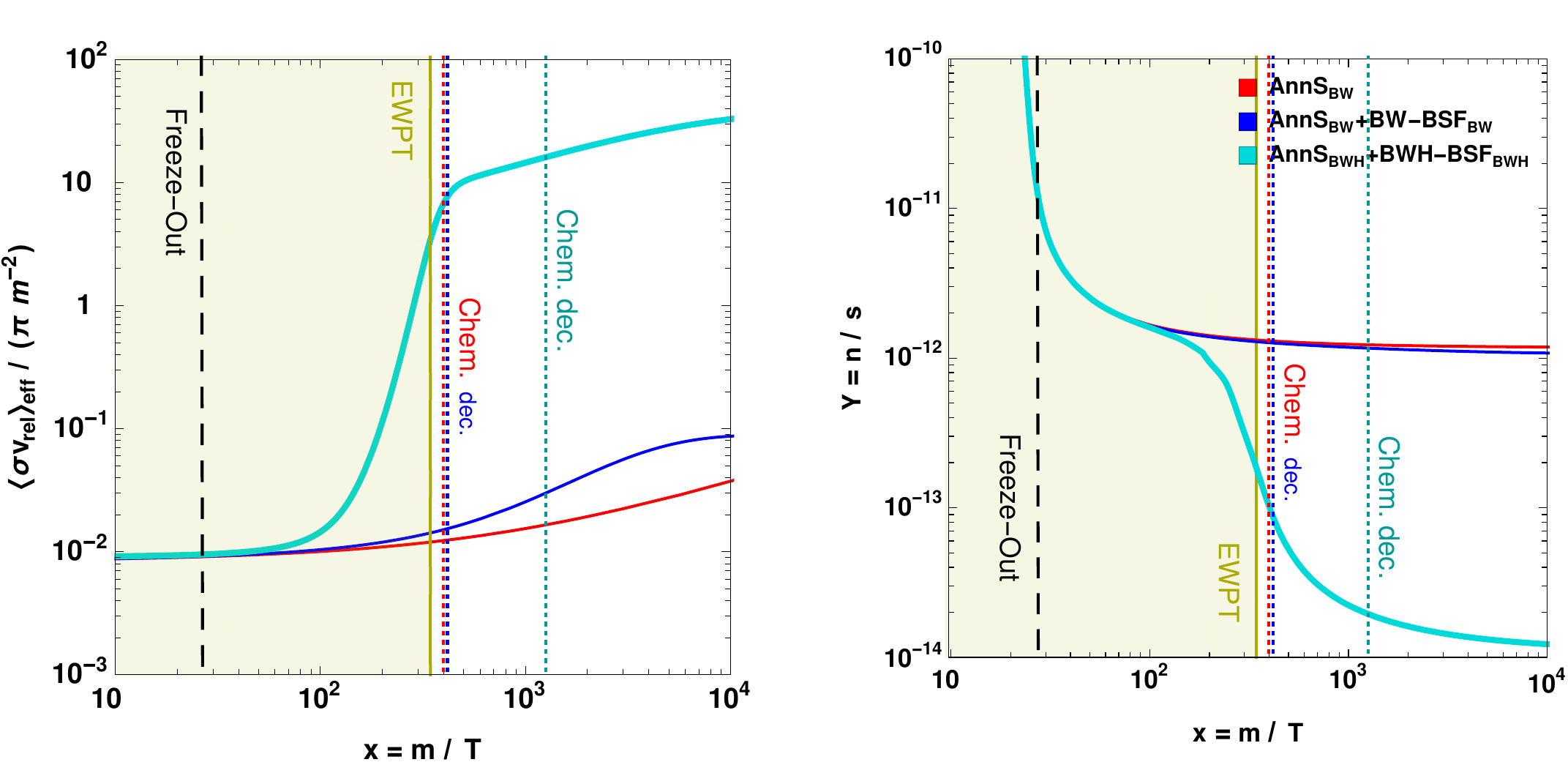}
\caption{\label{fig:Decoupling} 
The effective cross-section $\<\sigma\vrel\>_{\eff}/(\pi m^{-2})$ and the dark matter yield $Y\equiv n/s$, vs the time parameter $x = m/T$. We also mark the time of freeze-out, the EWPT, and the chemical decoupling for the three cases in the legend. We have used $m=50~\TeV$ and $\aH=0.2$.}
\bigskip
\includegraphics[width=0.49\textwidth]{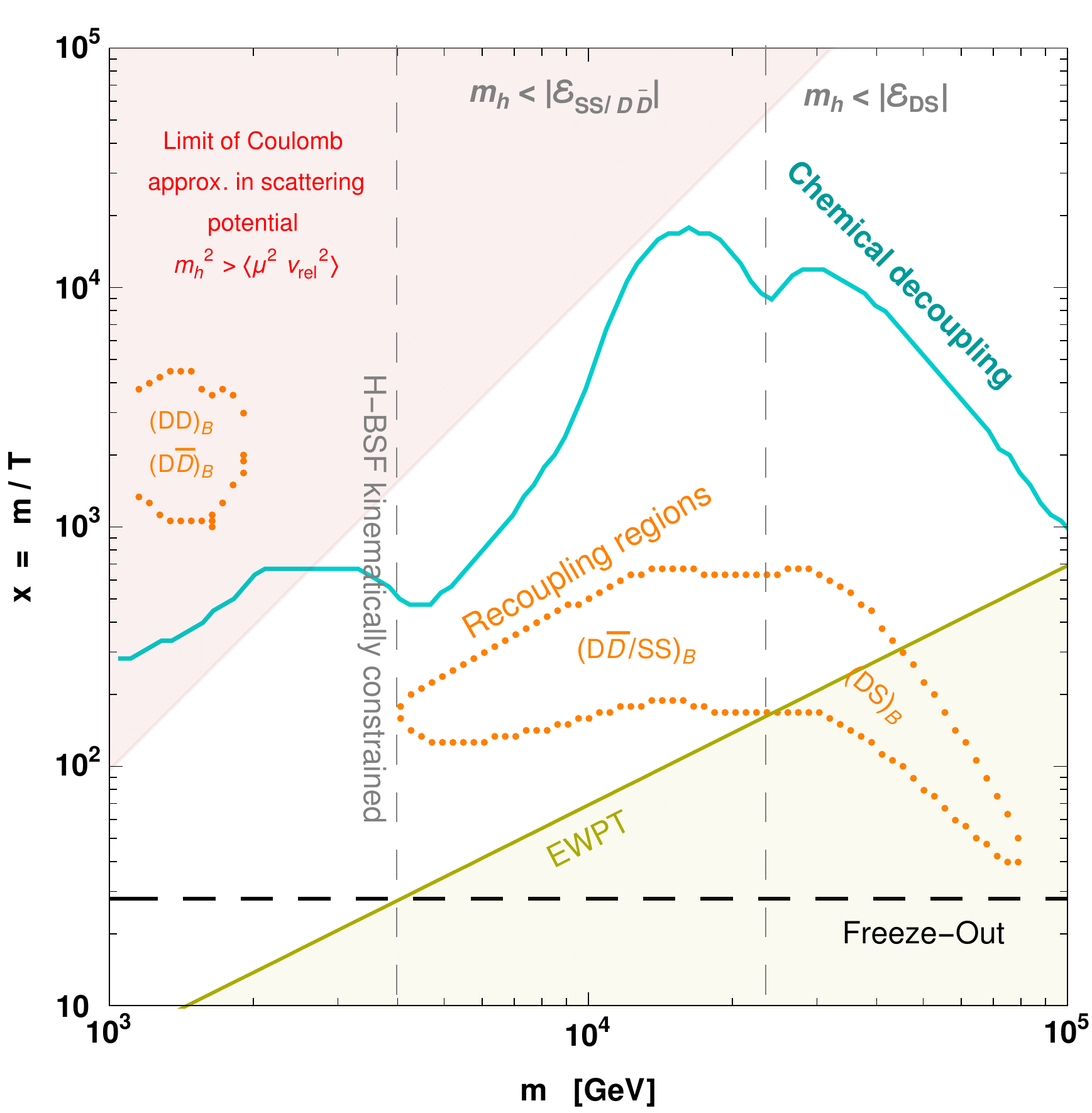}
\caption{\label{fig:Timeline} 
Various timeposts in the DM decoupling. See text for discussion. The coupling $\aH$ is chosen as function of the mass $m$ such that the observed DM density is reproduced by the full calculation (cf.~cyan line in \cref{fig:aHvsMass}.) For the recoupling intervals, we indicate the bound state that has the dominant effect.}
\end{figure}

In \cref{fig:Decoupling}, we present an example of the time evolution of the effective cross-section and the DM density. For the parameters chosen, the exponential increase of $\<\sigma \vrel\>_{\eff}$ due to BSF when the ionisation processes cease, gives rise to a second period of DM destruction that decreases the DM density by two orders of magnitude! 
In \cref{fig:Timeline}, we show the timeline of the DM thermal decoupling. We define the recoupling period of DM destruction due to BSF as the interval between the two occurrences when $d^2 (\ln Y) / d(\ln x)^2 =0$, and the chemical decoupling as the latest time when $d (\ln Y) / d(\ln x) =10\%$. 
In the same plot, we also mark the EWPT, as well as the time beyond which the finite Higgs mass affects its long-range effect. 
Since in part of the parameter space, the recoupling occurs after the EWPT and the full chemical decoupling occurs even later, the effect of BSF via Higgs emission is most important for the range of DM masses where the binding energies exceed the Higgs boson mass, $\mh \simeq 125~\GeV$. These ranges are also marked in \cref{fig:Timeline}. 
(We discuss the validity of various approximations, including that of electroweak symmetry, in \cref{sec:FreezeOut_Approximations}.)

\begin{figure}[t!]
\centering
\hfill
\includegraphics[height=0.49\textwidth]{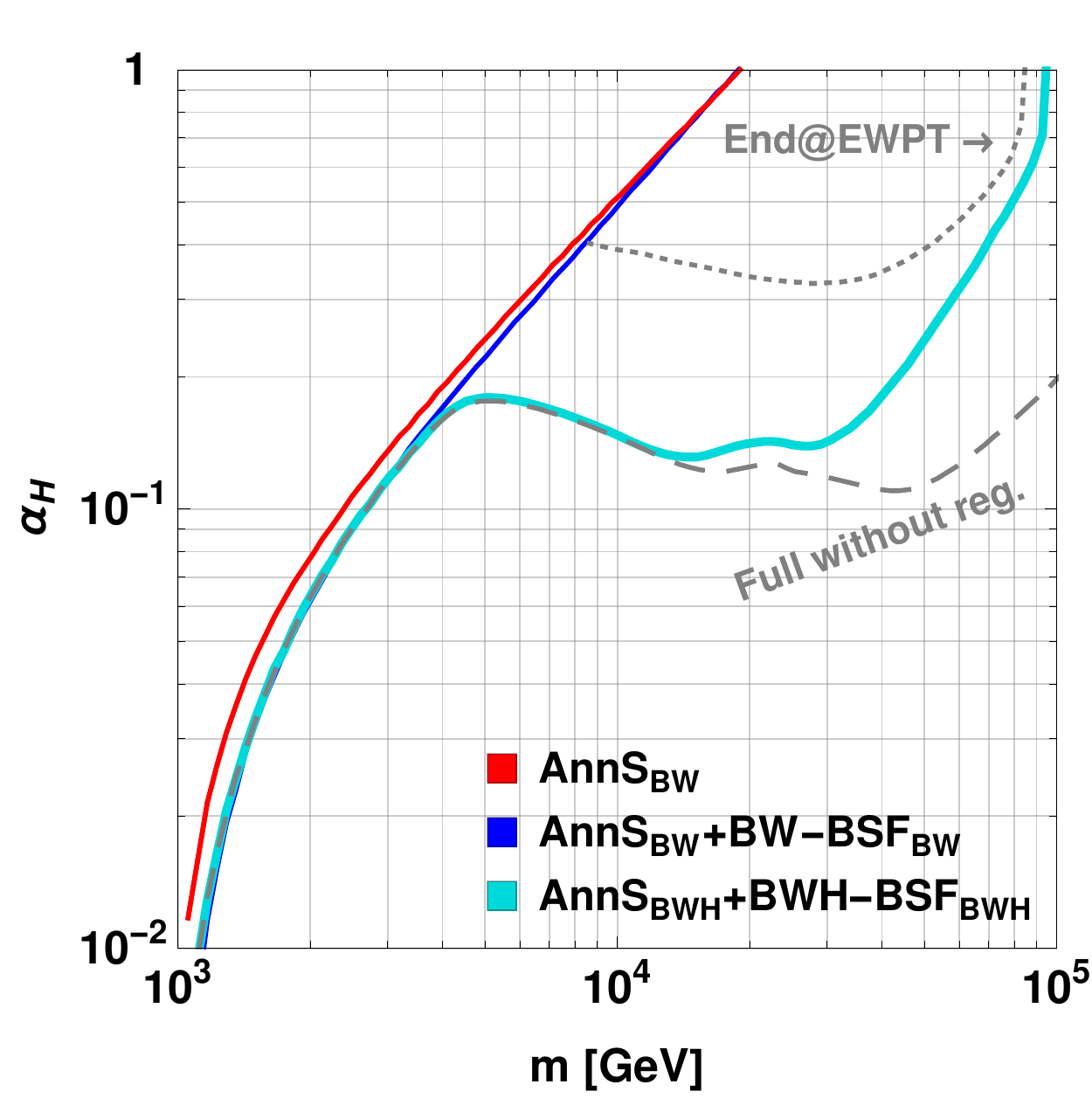}
\hfill
\includegraphics[height=0.49\textwidth]{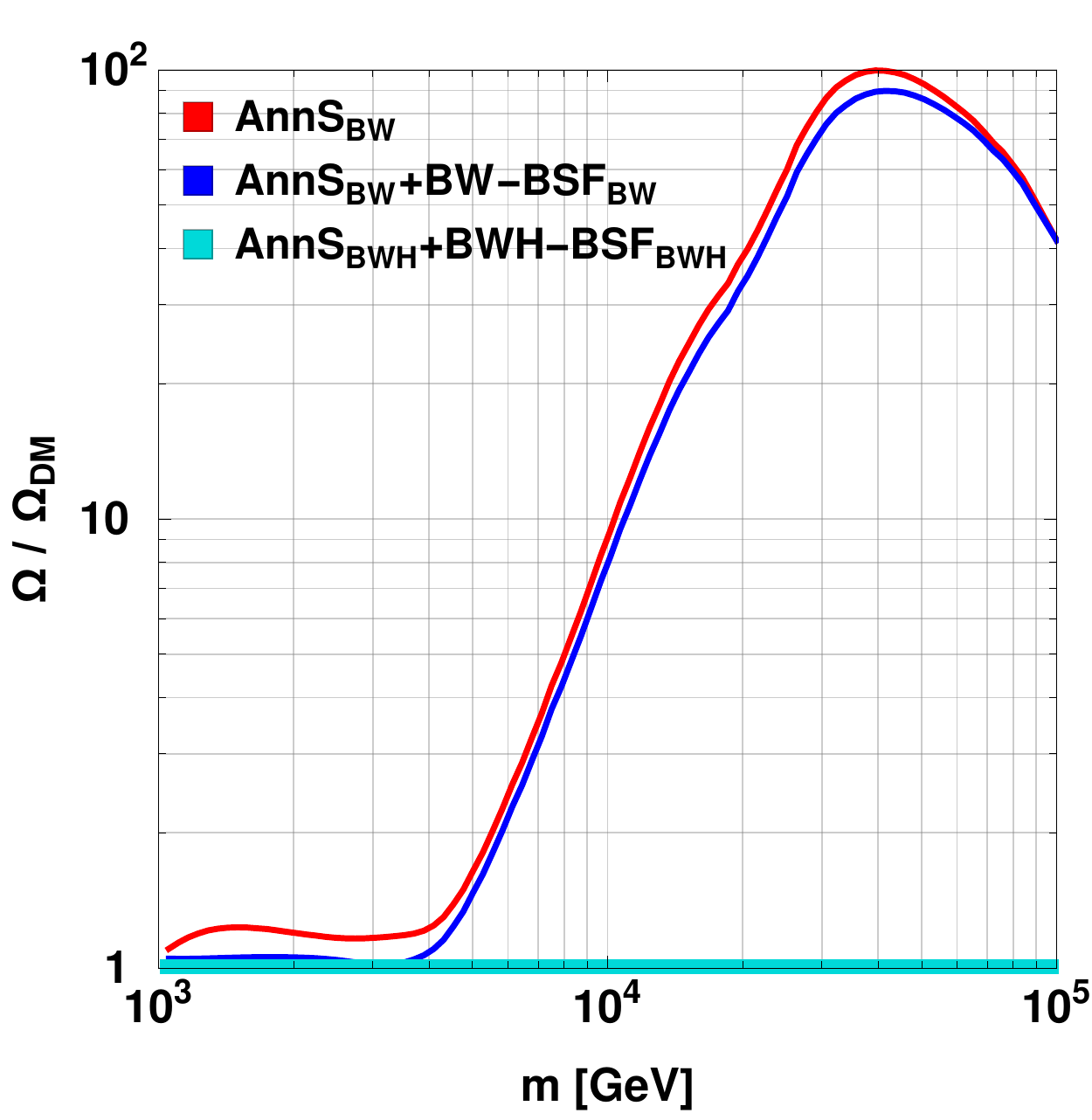}
\hfill
\caption{\label{fig:aHvsMass} 
\emph{Left:} $\aH$ vs $m$ such that the observed DM density is attained via thermal decoupling, when different combinations of effects are considered, as described in \cref{tab:CasesForComparison}. Note that the DM mass is $m_{\DM} = m - \sqrt{4\pi \aH} \vH$ with $\vH \simeq 246~\GeV$, and does not differ substantially from $m$ along any of the lines.
In grey lines, we show the result if the Boltzmann equations are integrated only up until the EWPT (\emph{dotted}), and if the cross-sections are \emph{not} regulated according to \cite[section~3.6]{Oncala:2021tkz} (\emph{dashed}.) 
\emph{Right:} The effect of the various processes on the relic density. For all lines, $\aH$ is determined as a function of $m$ by the full computation on the left (cyan line), but for each line here the Boltzmann equations include only the processes indicated in the legend.}

\bigskip\bigskip

\includegraphics[width=0.49\textwidth]{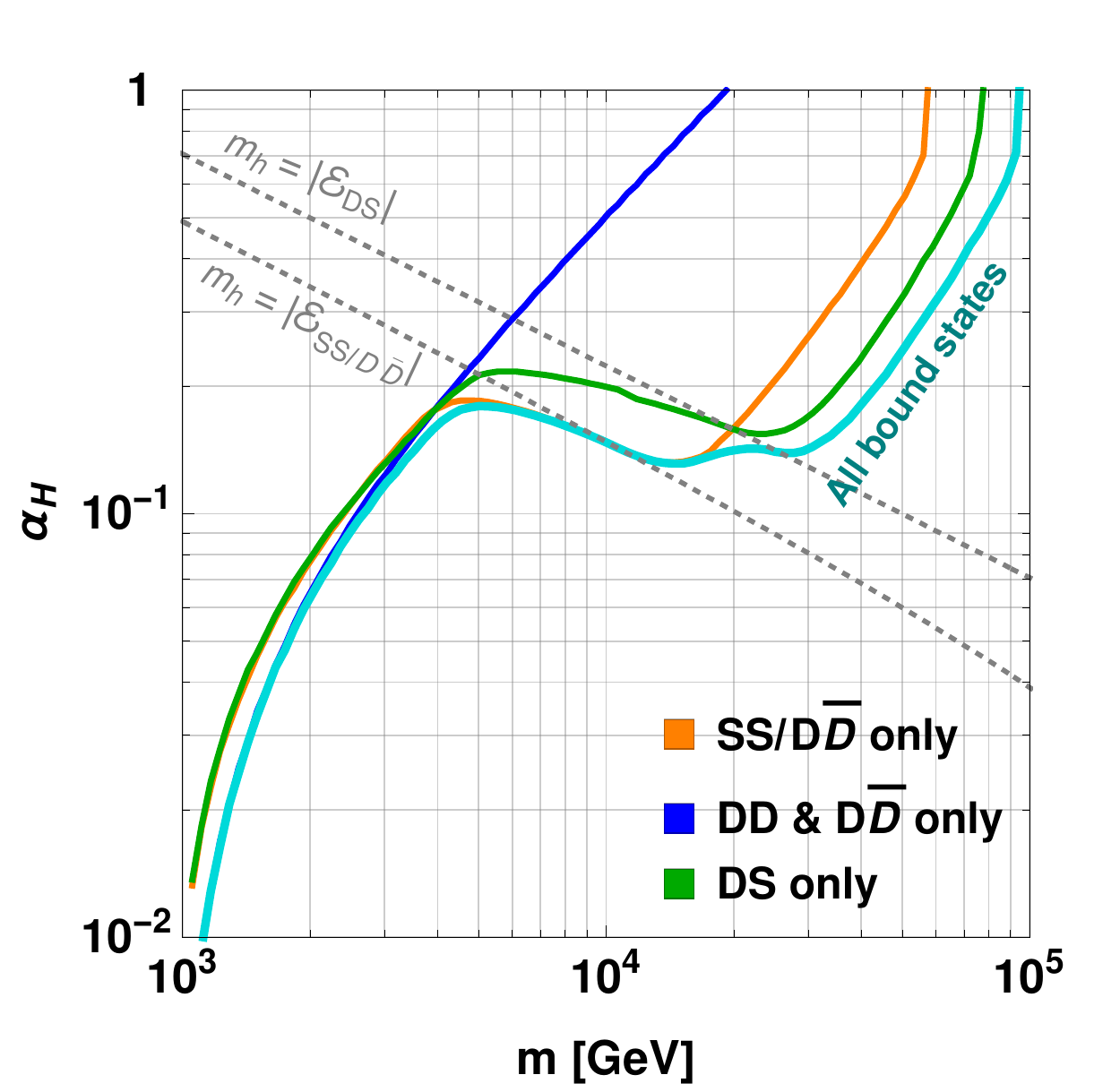}
\caption{\label{fig:aHvsMass_BoundStates} 
$\aH$ vs $m$ such that the observed DM density is attained via thermal decoupling, considering the contributions of the various bound states separately.}
\end{figure}

In~\cref{fig:aHvsMass}, we show the values of $\aH$ vs. $m$ that reproduce the observed DM density, as well as the impact of the various processes on the relic density. As already seen in \cref{fig:Decoupling}, at $m \gtrsim$~few~TeV, BSF via emission of a Higgs doublet is estimated to decrease the relic density by up to two orders of magnitude. The implications are twofold. For a fixed mass $m$, the coupling $\aH$ is predicted to be almost up to an order of magnitude smaller than when neglecting BSF via Higgs emission. This should be expected to change (relax) experimental constraints very significantly. Conversely, for a given coupling, a much larger $m$ is anticipated. In fact, DM masses almost up to the unitarity limit can be attained for $\aH < 1$. (We discuss the unitarity limit in more detail in \cref{sec:FreezeOut_Unitarity}.) This motivates experimental searches at very high masses.

To understand better the effect of the various bound states, in \cref{fig:aHvsMass_BoundStates} we show the $\aH-m$ relation determined by considering direct annihilation plus each of the four bound states separately.  The spin-1 $DD$ and $D\bar{D}$ bound states have only a small effect because their binding energy is independent of $\aH$ and somewhat small. This implies that ionisations inhibit the DM destruction via their formation until late, when BSF via Higgs emission is kinematically blocked, and BSF via $B$ or $W$ emission is not sufficiently fast to overcome the suppression due to the low DM density. Passing on to the $SS/D\bar{D}$ and $DS$ bound states, for the lower range of $m$ and $\aH$, their formation destroys DM efficiently after the EWPT. Thus the threshold for their effect being important is set by $|{\cal E}_{\Bcal}| > \mh \simeq 125~\GeV$, as the grey dotted lines in \cref{fig:aHvsMass_BoundStates} indicate.

\begin{figure}[h!]
\centering
\includegraphics[width=\textwidth]{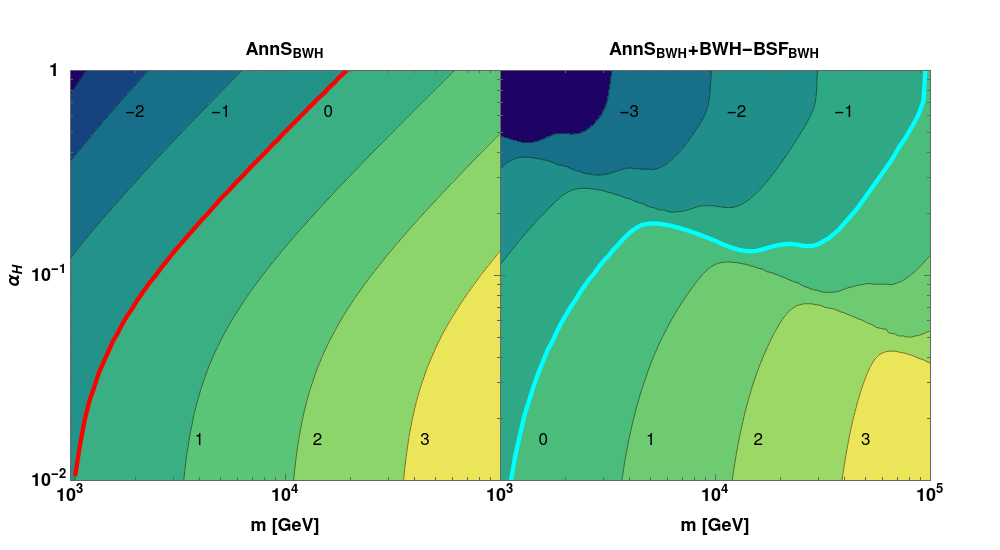}
\caption{\label{fig:OmegaContours} 
Contours of $\log_{10}(\Omega/\Omega_{\DM})$ (with the values indicated in black), when considering ${\rm AnnS}_{\BW}$ only (\emph{left}), and ${\rm AnnS}_{\BWH} + \BWH{\rm -BSF}_{\BWH}$ (\emph{right}.) The red and cyan lines mark $\Omega = \Omega_{\DM}$. Note that the DM mass, $m_{\DM} = m - \sqrt{4\pi \aH} \vH$, differs significantly from $m$ only at the top left corners of the plots.}
\end{figure}

Even away from the correlation of parameters that reproduces the observed DM density, the BSF effect on the relic abundance of the stable species can be very large as seen in \cref{fig:OmegaContours}. The parameter space where the relic density is cosmologically insignificant is greatly enlarged. This is important for scenarios that do not aspire to explain the DM density, but nevertheless predict the existence of stable particles.

\clearpage
\subsection{Major approximations and their validity \label{sec:FreezeOut_Approximations}}

We now summarise the main approximations made in our analysis and comment on their potential effect on the estimated relic density.
\begin{enumerate}[(i)]
\item \label{item:Approx_GroundStatesOnly}
\emph{Considered only ground-level bound states.} 

BSF via vector or neutral scalar emission is dominated by dipole and quadrapole moments respectively. In these cases, the capture into the ground state is the dominant BSF process~\cite{vonHarling:2014kha,Petraki:2015hla,Petraki:2016cnz,Harz:2018csl,Oncala:2018bvl}, the reason being twofold: it is the most exothermic process, and the overlap of scattering and bound state wavefunctions is larger. 

In contrast, BSF via emission of a charged scalar is a monopole transition, and the capture into excited states can be comparable to or faster than the capture into the ground state, despite the latter being more exothermic~\cite{Oncala:2019yvj}. This suggests that in the present model, capture into excited states via Higgs emission may be important. 

Nevertheless, independently of the BSF cross-sections, the relative effect of the excited states on the relic density is moderated by their smaller binding energy that renders their ionisation efficient until later. 
We thus anticipate that in the present model excited states may have a significant albeit not dominant effect that would further diminish the relic density and alter the coupling-mass relation along the direction found here. This is worth pursuing in more detail in the future.

\item \label{item:Approx_UnitarityPrescription}
\emph{Regularisation of inelastic cross-sections in parametric regimes where BSF via Higgs emission approaches or appears to exceed the unitarity limit.}

The efficiency of BSF via emission of a charged scalar implies that the computed cross-sections may reach or violate the upper limit on inelastic cross-sections implied by unitarity (cf.~\cref{eq:UnitarityLimit_sigmav}) even for small or moderate values of the coupling of the interacting particles to the scalar~\cite{Oncala:2019yvj}. The restoration of unitarity implies that resummed higher order corrections, i.e.~higher-order contributions to the non-relativistic potential, must be considered~\cite{Baldes:2017gzw}. 
In the context of the present model, the problem was discussed in ref.~\cite[section~3.6]{Oncala:2021tkz}, where the regularisation scheme of ref.~\cite{Blum:2016nrz} was adopted as an effective method to ensure that unitarity is not violated. However, as discussed in ref.~\cite{Oncala:2021tkz}, this scheme is not entirely suitable for the case of interest.

In \cref{fig:aHvsMass} we compare the $\aH-m$ relations with and without regularisation. Clearly, at large $m$ the effect is significant; the regularisation of the cross-sections ensures that $m$ does not exceed the unitarity limit on the mass of thermal relic DM~\cite{Griest:1989wd,Baldes:2017gzw}, which we discuss in \cref{sec:FreezeOut_Unitarity}. This suggests that working out a more accurate regularisation scheme that would address the issues discussed in  ref.~\cite{Oncala:2021tkz} may be important in order to obtain more accurate results. We leave this for future work.

\item \label{item:Approx_HiggsCoulomb}
\emph{Neglected the Higgs mass in the Higgs-mediated potential.} 

The validity of the Coulomb approximation for the Higgs-mediated potential has been discussed in \cite[section~2.3.3]{Oncala:2021tkz}, where the relevant conditions have been put forward. Here we discuss their validity during the DM thermal decoupling. 

Scattering states: In a thermal bath, the condition  $\mu \vrel > \mH$ for the validity of the Coulomb approximation implies $\sqrt{3m T/2} \gtrsim \mH$. Considering the Higgs doublet mass \eqref{eq:mH} before the EWPT, this becomes $T \lesssim 3m$, which clearly covers all of the range of interest for the DM freeze-out ($T\lesssim m/25$.) Below the EWPT, where $\mH \to \mh \simeq 125~\GeV$, the condition is satisfied until after the DM chemical decoupling, as shown in \cref{fig:Timeline}. Therefore, the Coulomb approximation does not pose any problem.

Bound states: The condition $\mu \aH > {\rm few} \times \mH$, becomes $x \aH > {\rm few}$ before the EWPT, with $x=m/T$. This is satisfied for all relevant $x$ and $\aH$ for which BSF has an effect ($x\gtrsim 50$ and $\aH \gtrsim 0.1$, cf.~\cref{fig:aHvsMass,fig:Timeline}.) It is easy to check that this condition is also satisfied below the EWPT for all relevant DM masses and couplings ($m > 5~\TeV$ and $\aH \gtrsim 0.1$, cf.~\cref{fig:aHvsMass}.)
Note that this is not coincidental; BSF via Higgs emission does not have a significant effect for lower $\aH$ values because of the phase-space suppression due to the Higgs mass (cf.~\cref{fig:aHvsMass_BoundStates}.) The estimation here thus confirms the argument of ref.~\cite{Oncala:2021tkz} that bound states are nearly Coulombic in the parameter space where their formation is kinematically allowed and significant.

\item \label{item:Approx_EWsymmetry}
\emph{Assumed electroweak symmetry.}

In \cref{fig:Timeline}, we see that the DM destruction via BSF may be efficient after the EWPT. The breaking of the electroweak symmetry has several important implications that we now discuss.
\begin{enumerate}[a.]
\item 
\emph{The Goldstone modes of the Higgs doublet are absorbed by the $Z,W^\pm$ bosons.}

\begin{itemize}
\item 
BSF via emission of a Higgs doublet in the unbroken electroweak phase corresponds to BSF via emission of $h$ or the longitudinal modes of the $Z,W^\pm$ bosons in the broken electroweak phase. The Goldstone boson equivalence theorem implies that the amplitudes for BSF via emission of the longitudinal $Z, W^{\pm}$ components are the same as those for the corresponding processes in the unbroken electroweak phase, in the limit that the energy of the emitted vector boson is much larger than its mass, $m_{\mathsmaller{Z,W}}^2 / \omega^2 \ll 1$. 

In our computation, the phase-space suppression sets $\mH \to \mh \simeq 125~\GeV$ after the EWPT, ensuring that $\mh /\omega <1$ or equivalently $m_{\mathsmaller{Z,W}}^2 / \omega^2 < 0.5$. We thus regard the approximation as acceptable, especially in the parameter space away from the phase-space thresholds (cf.~\cref{fig:aHvsMass_BoundStates}.)  

The importance of monopole BSF processes in a broken gauge phase due to the Goldstone boson equivalence theorem was previously pointed out in ref.~\cite{Ko:2019wxq}.

\item 
The potential mediated by the Higgs doublet in the unbroken electroweak phase is mediated by $h$ and the longitudinal $Z,W^{\pm}$ components in the broken phase.  To compute the non-relativistic potential generated by the latter, we need their contribution to the vector boson propagators,
\begin{align}
\dfrac{\im}{q^2 - \mV^2} \dfrac{q^\mu q^\nu}{\mV^2} ,
\label{eq:LongitudinalProp}
\end{align}
where $q$ and $\mV = \gV \vH/2$ denote the vector boson momentum and mass, for $V = Z,W^{\pm}$, with $\gZ = \sqrt{g_1^2+g_2^2}$ and $\gW = g_2$. 
In general, the exchange of $Z,W^{\pm}$ between a pair of $\mathbb{Z}_2$-odd particles may change the mass eigenstate on each leg. (Indeed, in the model under consideration, the $Z, W^\pm$ bosons couple only non-diagonally to the mass eigenstates, cf.~\cref{eq:Lagrangian_MassNeutral}.) Considering \eqref{eq:LongitudinalProp}, the contribution from the exchange of the longitudinal $Z,W^{\pm}$ components to the 2PI kernels (cf.~ref.~\cite[section~2]{Oncala:2021tkz}) is proportional to
\begin{align}
{\cal K}_{\mathsmaller{L}} 
&\propto 
[\bar{u}(p_1') \im \gV \slashed{q} u (p_1) ] 
[\bar{u}(p_2') \im \gV \slashed{q} u (p_2) ] / \mV^2
\nn \\
&=
(\im \gV)^2 
(m_1'-m_1) \bar{u}(p_1') u (p_1) \
(m_2-m_2') \bar{u}(p_2') u (p_2) / \mV^2 ,
\label{eq:KL0}
\end{align}
where  $q=p_1'-p_1 = p_2 - p_2'$, and we used the Dirac equation 
$\slashed{p}u(p) = mu(p)$. Considering the mass splittings $\sim y\vH$, this becomes
\begin{align}
{\cal K}_{\mathsmaller{L}} 
&\propto 
\gV^2 (y\vH)^2 (2m)^2 / \mV^2 \propto y^2 m^2 .
\label{eq:KL}
\end{align} 
\Cref{eq:KL} shows that the potentials generated by the exchange of the longitudinal $Z,W^{\pm}$ is indeed proportional to the coupling to the Higgs doublet. The range of the potentials are $\mV^{-1} > \mh^{-1}$, thus the arguments presented in item~\eqref{item:Approx_HiggsCoulomb} for the Coulomb approximation remain valid. An analogous result has been obtained in \cite{Ko:2019wxq} for a broken $U(1)$ model.

Note that in \cref{eq:KL0,eq:KL} we omitted various numerical factors and signs for simplicity, and focused on deriving the scaling of the 2PI kernel. Considering these factors in detail reproduces the Higgs-doublet mediated potential (aside from the screening scale.) 

\end{itemize}

\item \emph{The Weak gauge bosons become massive.}

The non-zero $Z,W^\pm$ masses curtail the range of the potentials generated by the exchange of both their transverse and longitudinal components, and introduce phase-space suppression to the BSF processes occurring via their emission. The validity of the Coulomb approximation for the $Z,W^\pm$ bosons can be assessed as in the preceding discussion for the Higgs. However, in the present model, the $B,W$-generated potentials and BSF via $B$ or $W$ emission do not have a significant effect, due to the fact that one of the dark multiplets is a gauge singlet and the other belongs to a small representation. We thus do not consider the transverse $Z,W^\pm$ components further. The effect of the longitudinal $Z,W^\pm$ components was discussed above.

\item \emph{The components of the DM multiplets acquire different masses.} 

After acquiring a mass splitting, the various pairs of $\mathbb{Z}_2$-odd particles can oscillate into each other according to the non-relativistic potentials computed in \cite[section~2]{Oncala:2021tkz} provided that the kinetic energy of their relative motion exceeds their mass difference. This necessitates $m \vrel^2/4 > 2 y \vH$, which, upon thermal averaging, becomes $T> (4/3) \sqrt{4\pi \aH} \vH$. This condition is \emph{not} satisfied below the EWPT for the $\aH$ values of interest ($\aH \gtrsim 0.1$.) We thus expect that the rates of some of the processes below the EWPT will be lower than estimated here. 

This is probably the most severe limitation of our computation. To assess its impact, in \cref{fig:aHvsMass} we include the coupling-mass relation obtained by integrating the Boltzmann equations only up to the EWPT. Clearly, a proper treatment would result in an $\aH-m$ relation between our this and the result obtained by integrating until late times. 
We see that even when the integration stops at the EWPT, the Higgs effect is still very significant, even if it appears only for larger $\aH$ values. The impact on the relic density reaches up to a factor of a few.

\end{enumerate}  
\end{enumerate}

\subsection{Unitarity limit on the dark matter mass \label{sec:FreezeOut_Unitarity}}

The unitarity of the $S$ matrix sets an upper limit on the partial-wave inelastic cross-sections, 
\begin{align}
\sigma_{\ell}^{\rm inel} 
\leqslant \sigma^{\rm uni}_{\ell}
= \dfrac{(2\ell +1)\pi}{k^2}
\simeq \dfrac{(2\ell+1) \pi}{\mu^2 \vrel^2},
\label{eq:UnitarityLimit_sigmav}
\end{align}
where $\ell$ is the partial wave and $k$ is the momentum of either of the interacting particles in the CM frame. The last approximation in \cref{eq:UnitarityLimit_sigmav} concerns the non-relativistic regime, where $k=\mu \vrel$ with $\mu$ being the reduced mass.

The upper limit \eqref{eq:UnitarityLimit_sigmav} suggests that for very large masses, annihilations in the early universe may not suffice to reduce the density of thermalised particles to the observed DM value. It thus sets an upper bound on the mass of thermal relic DM annihilating predominantly via a finite number of partial waves in the early universe~\cite{Griest:1989wd}. For self-conjugate DM in thermal equilibrium with the SM plasma, this is~\cite{vonHarling:2014kha,Baldes:2017gzw}\footnote{If DM annihilates into a dark plasma that has different temperature than the SM plasma or includes many relativistic dof during DM freeze-out, then this value may somewhat change. Moreover, departures from thermal cosmology, such as episodes of entropy injection (see e.g.~\cite{Cirelli:2018iax}), imply that larger $m_{\DM}$ values may be permissible.} 
\begin{align}
m_{\DM,\ell} \lesssim 197~\TeV \times \left \{
\begin{array}{ll}
\sqrt{2\ell+1}, &~~\text{solely}~\ell, \\
\ell+1, &~~0 \leqslant \ell \leqslant \ell_{\max}.
\end{array}
\right.
\label{eq:UnitarityLimit_mDM}
\end{align}
\Cref{eq:UnitarityLimit_mDM} is modified by $1/\sqrt{2}$ in the case of non-self-conjugate DM.

The parametric dependence of $\sigma^{\rm uni}_{\ell}$ on $\mu$ and $\vrel$ implies that the limit~\eqref{eq:UnitarityLimit_sigmav} can be attained down to arbitrarily low velocities --- thus the upper limit \eqref{eq:UnitarityLimit_mDM} on the mass of thermal-relic DM can be reached --- only if there is an attractive long-range force between the interacting particles, and provided of course that the relevant couplings are sufficiently large~\cite{Baldes:2017gzw}. Attractive long-range interactions imply also the existence of bound states, whose formation and subsequent decay may decrease the DM abundance more efficiently than direct annihilation~\cite{vonHarling:2014kha}. This means that BSF may essentially be the dominant process that saturates the unitarity limit \eqref{eq:UnitarityLimit_sigmav}, and that additional partial waves to those dominating in the annihilation processes may become important, thereby increasing the upper limit on the DM mass~\cite{Baldes:2017gzw}.

In general, the Weak interactions of the Standard Model are not sufficiently strong to generate cross-sections that approach the unitarity limit~\eqref{eq:UnitarityLimit_sigmav}, unless perhaps the interacting particles belong to very large $\SUL$ representations. However, BSF via emission of a scalar charged under a symmetry can be very efficient even for small couplings~\cite{Oncala:2019yvj}. Here, we have seen that BSF via Higgs emission can raise the predicted WIMP mass very significantly, bringing it potentially close to the unitarity limit.

\clearpage
\section{Conclusion \label{Sec:Conclusion}}

Our DM searches are currently at the onset of the exploration of the multi-TeV regime with a variety of existing and upcoming telescopes observing high-energy cosmic rays. In this mass regime, within the thermal-relic scenario, the DM interactions are expected to manifest as long-range and give rise to non-perturbative effects, in particular the Sommerfeld effect and the formation of bound states. These effects may operate in the early universe during the DM thermal decoupling, as well as inside DM haloes today, and significantly alter the DM phenomenology.

In the present work, consisting of this and a companion paper~\cite{Oncala:2021tkz}, we have considered the role of the Higgs doublet as a light force mediator, in the thermal decoupling of multi-TeV WIMP DM. We have shown that the Higgs-doublet-mediated potential between DM particles and the formation of DM bound states via Higgs-doublet emission can dramatically change the predicted relic density. This, in turn, alters the coupling-mass relation that reproduces the observed DM abundance. Moreover, it greatly expands the parameter space where the stable relics do not overclose the universe, even if they are a subdominant component of DM. In the former case, the modified coupling-mass relation implies that on one hand, for a given DM mass, existing constraints may be significantly relaxed, and on the other hand, DM may be much heavier than previously anticipated, potentially approaching the unitarity limit.

While the amplitude for BSF via Higgs-doublet emission can be quite large even for small couplings of the DM multiplets to the Higgs, the Higgs-doublet mass introduces a kinematic suppression to the cross-section that renders this effect relevant for larger DM masses and/or couplings to the Higgs. In the specific singlet-doublet scenario considered here, we found that the effect is significant for $m\gtrsim 5~\TeV$ and $\aH \gtrsim 0.1$. 
However, in models involving larger $\SUL$ representations, the gauge interactions contribute more significantly to the binding energy of the bound states, thereby rendering the phase-space suppression less significant. We thus expect that the effect on the relic density will be important even for lower couplings. 

Finally, we note that the capture into excited bound levels, which we neglected here, may also have a sizeable effect due to the monopole nature of the transitions occurring via Higgs-doublet emission. 
On the other hand, we have found that including BSF through scattering on the relativistic thermal bath via an off-shell Higgs doublet does not affect the relic density significantly.

\clearpage
\section*{Acknowledgements}
This work was supported by the ANR ACHN 2015 grant (``TheIntricateDark" project), and by the NWO Vidi grant ``Self-interacting asymmetric dark matter".

\bibliography{Bibliography.bib}

\end{document}